\documentclass[prd,aps,showpacs,,nofootinbib,superscriptaddress,preprintnumbers,epsf,psf]{revtex4}
\usepackage[english]{babel}
\usepackage{pstricks}
\usepackage{axodraw}
\usepackage[dvips]{graphicx}
\usepackage{epsf}

\newcommand{\AmS}{{\protect\the\textfont2
  A\kern-.1667em\lower.5ex\hbox{M}\kern-.125emS}}
\newcommand{\tr}{\,{\rm Tr}}
\begin{document}
\makeatletter
 \def\preprint#1{% \def\@preprint{\noindent\hfill\hbox{#1}\vskip 10pt}%
 }
\preprint{\begin{tabular}{l}
      arXiv \\
      CPHT
    \end{tabular}
 }
%
% \draft command makes pacs numbers print
%\draft
%
\title{NLO corrections to timelike, spacelike and double deeply virtual Compton scattering.}
\author{B. Pire}
 \address{ CPHT, {\'E}cole Polytechnique, CNRS, 91128 Palaiseau, France}
  \author{     L. Szymanowski$^2$ 
        and
J. Wagner}
 \address{ Soltan Institute for Nuclear Studies, Ho\.{z}a 69, 00-681
Warsaw, Poland}

\begin{abstract}
We calculate the $O (\alpha_s)$ corrections to the timelike, spacelike and
double deeply virtual Compton scattering amplitudes in the generalized
Bj\"orken scaling region. Special attention is devoted to studies of the
difference between the next to leading order timelike and spacelike   
coefficient functions, which plays for this process a role analogous to 
the large K factor which was much discussed in the analysis of inclusive Drell Yan cross sections. 
Also in the present studies the timelike nature of the hard scale gives rise to new absorptive part of the
amplitude and to the presence of characteristic $\pi^2$ terms, which can potentially lead to sizeable corrections. 
\end{abstract}
%
% insert suggested PACS numbers in braces on next line
\maketitle
\narrowtext
%%%%%%%%%%%%%%%%%%%%%%%%%%%%%%%%%%%%%%%%%%%%%%%%%%%%%%%%%%%%%%%%%%%%%%%%%%%%
\noindent
\section{Introduction.}
Data on deeply virtual Compton scattering (DVCS) are now available from various experimental settings \cite{DVCSexp} and different strategies are proposed \cite{fitting} to extract from them the physical knowledge on nucleon structure encoded in generalized parton distributions (GPD) \cite{historyofDVCS,gpdrev}. These attempts are usually based on a leading order QCD analysis, although the importance of next order terms has often been emphasized, in particular with respect to the dangerous factorization scale choice dependence \cite{Anikin}. Historically, one can note that the  understanding of inclusive reactions (Drell Yan reactions, large $p_T$ hadron or jet production) in the framework of collinear QCD factorization has waited for an analysis including next-to-leading order (NLO) (or even next-to-next-to-leading order) corrections. Indeed, complete NLO calculations \cite{JiO,Mankiewicz,Belitsky:1997rh,Belitsky:1999sg} are available for the DVCS reaction and there is no indication that they are negligible in the kinematics relevant for current or near future experiments.
Deeply virtual Compton scattering is only one case of the general double DVCS(DDVCS) reaction
\begin{equation}
\gamma^*(q_{in}) N \to \gamma^*(q_{out}) N'\,,
\end{equation}
where the final photon is on shell, $q_{out}^2 =0$. The converse case where $q^2_{in} = 0$, often called timelike Compton scattering (TCS), has been theoretically discussed at medium \cite{BDP} and very large \cite{PSW} energy and first data are being analyzed \cite{NadelTuronski:2009zz}. The double DVCS case has been discussed in Ref. \cite{DDVCS}.

It has been shown that the understanding of DVCS data needs higher order calculations for a reasonable extraction of GPDs to be possible \cite{NLOphenoDVCS}. This is likely to be even more the case for TCS. Indeed, TCS and DVCS amplitudes are identical (up to a complex conjugation) at lowest order in $\alpha_S$ but differ at next to leading order, in particular because of the quite different analytic structure of these reactions. Indeed the production of a timelike photon enables the production of intermediate states in some channels which were kinematically forbidden in he DVCS case. This opens the way to new absorptive parts of the amplitude.  Soon, experiments will be performed at JLab at 12 GeV which will enable to test the universality of GPDs extracted from DVCS and from TCS, provided NLO corrections are taken into account. Experiments at higher energies, e.g. in ultraperipheral collisions at RHIC and LHC, may even become sensitive to gluon GPDs which enter the amplitude only at NLO level.

Former experience with inclusive deep reactions teaches us that NLO corrections are likely to be more important in timelike reactions than in corresponding spacelike ones. The well-known example of the Drell Yan K-factor teaches us that NLO corrections are sizeable in timelike processes, because of $i \pi$  factors coming from $log(-Q^2/\mu_F^2)$ terms which often exponentiate when soft gluon resummation is taken care of \cite{DIS,Kfactor}.

The results for TCS should be indicative of other exclusive reactions with a timelike scale, as $\pi N \to l^+l^-N'$ discussed in \cite{BDPpi} which may be accessed in the Compass experiment at CERN or at J-Parc, $e^+ e^- \to \gamma \pi \pi $ discussed in \cite{TGDA} to be compared  to $\gamma^* \gamma  \to \pi \pi $ analyzed in \cite{GDA}, or $\gamma^* N  \to N' \pi $\cite{TDA} to be compared  to $\bar N N' \to \gamma^*  \pi $ analyzed in \cite{TTDA}.

Our calculations	 are performed along the lines of Ref.\cite{JiO} (see also \cite{Mankiewicz}). Those earlier results where obtained in an unphysical region of parameters space and then by analytical continuation (due to simple analytical structure of hard DVCS amplitude) extrapolated to the physical region of DVCS. Not restricting the parameters enables us to get the  full result for the general kinematics (including TCS, DVCS and DDVCS). In earlier analysis the factorization scale  $\mu_F$ was kept equal to the hard scale $Q^2$. In our calculation they are independent, so one can check factorization  scale dependence. We calculate only the symmetric part of the amplitude which is dominant for the phenomenological analysis, as the main features of scattering amplitudes of DDVCS, DVCS and TCS are already clearly seen. We simplify the kinematics by restricting ourselves to the forward ($t=t_{min}$) region.
We leave the phenomenological analysis of our results to a future publication.
\section{Preliminaries.}
%%%%%%%%%%%%%%%%%%%%%%%%%%%%%%%%% FIGURE
\begin{figure}[hb!]
\begin{center}
\epsfxsize=0.5\textwidth
\epsffile{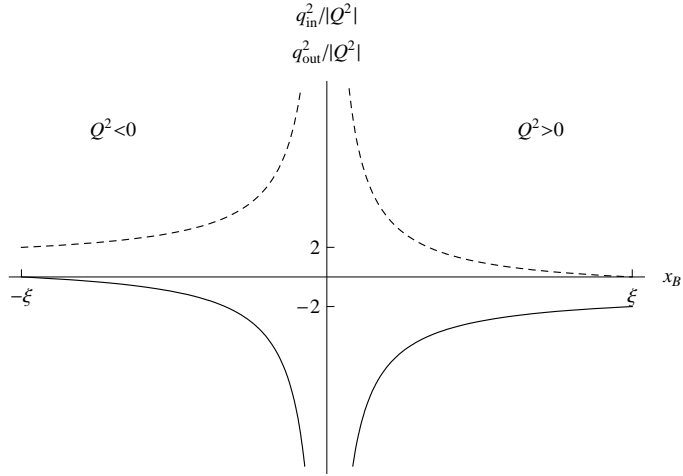}
\caption{\label{Fig:q1q2plot} Incoming and outgoing photon virtualities as a function of $x_B$. To get the physically interesting case in which incoming photon (solid line) is spacelike and outgoing photon (dashed line) is timelike, one has to choose $Q^2>0$ for $x_B>0$ and $Q^2<0$ for $x_B<0$. DVCS kinematics corresponds to $x_B =\xi$, and TCS to $x_B =-\xi$.}
\end{center}
\end{figure}
%%%%%%%%%%%%%%%%%%%%%%%%%%%%%%%%%%%%
As in \cite{JiO} we describe the kinematics of general Compton scattering in a symmetric way i.e. in the Bjorken limit, and in the forward limit ($q^\bot_{out}=0,P_T'=0$) momenta are assigned as follows: incoming photon $q_{in}=(q-\xi p)$, outgoing photon $(q_{out} = q+\xi p)$, incoming proton $P=(1+\xi)p$ and outgoing proton $P'=(1-\xi)p$, where:
\begin{eqnarray}
p &=& p^+ (1,0,0,1)\,,\nonumber \\
n &=& \frac{1}{2p^+} (1,0,0,-1)\,, \nonumber\\
q &=& -x_B p + \frac{Q^2}{2x_B}n\,, 
\label{eq:vecdef}
\end{eqnarray}
so: $pn =1$, $s = (p+q)^2 = \frac{1-x_B}{x_B}Q^2$ and $x_B = \frac{Q^2}{s+Q^2}$. We use the following vector decomposition $k^\sigma = k^+p^\sigma + k^-n^\sigma + k_\bot^\sigma$. With above definitions we arrive at the following equations for incoming and outgoing photon virtualities:
\begin{eqnarray}
q_{in}^2&=& -Q^2(1+\frac{\xi}{x_B}) \,,\nonumber \\
q_{out}^2&=&  -Q^2(1-\frac{\xi}{x_B})\,.
\label{eq:qinqout}
\end{eqnarray}
From this general kinematics we can get as a limit some physically interesting cases.  It is easy to check that to get incoming photon momentum spacelike and outgoing photon timelike, one has to choose $Q^2>0$ for $x_B>0$ and $Q^2<0$ for $x_B<0$. Deeply virtual Compton scattering is restored for $x_B = \xi$ and $Q^2>0$, timelike Compton scattering for $x_B = -\xi$ and $Q^2=-{Q'}^2<0$, and double deeply virtual Compton scattering for $0 < x_B < \xi$ and $Q^2>0$ or $0 > x_B  > -\xi$ and $Q^2<0$. This is illustrated by Fig. \ref{Fig:q1q2plot} which shows incoming and outgoing photon virtualities as a function of $x_B$.

We perform our calculation in $\overline{\textrm{MS}}$ scheme, with $D=4+\epsilon$ regularizing infrared
divergences, as all ultraviolet divergences cancel out. In the following we shall study only the symmetric part
of the full Compton scattering amplitude since it is phenomenologically the dominant part. Its factorized form reads:
\begin{eqnarray}
\mathcal{A}^{\mu\nu} = g_T^{\mu\nu}\int_{-1}^1 dx 
\left[
\sum_q^{n_F} \tilde{T}^q(x)\tilde{F}^q(x)+\tilde{T}^g(x)\tilde{F}^g(x)
\right] \,.
\label{eq:factorization}
\end{eqnarray}
Renormalized GPD's are defined as in \cite{JiO}, by:\\
\begin{eqnarray}
F^q(x,\xi)&=& \frac{1}{2} \int \frac{d \lambda}{2\pi}e^{-i\lambda x} 
 \bigg\langle  ~P'\bigg| \bar \psi_q  \left( \frac{\lambda}{2} n\right)\gamma^\mu  \psi_q \left( - \frac{\lambda}{2}n \right)\bigg | P ~\bigg\rangle n_\mu \,,
\nonumber \\
F^g(x,\xi)&=& - \frac{1}{2x} \int \frac{d \lambda}{2\pi}e^{-i\lambda x} 
~\bigg\langle P' \bigg|  F^{\mu \alpha}_a  \left( \frac{\lambda}{2}n \right)  F^{\nu}_{a \alpha} \left( - \frac{\lambda}{2} n\right)\bigg| P ~\bigg\rangle  n_\mu n_\nu \,.
\label{eq:GPD}
\end{eqnarray}

The connection between the bare quantities $\tilde{F}_q$, $\tilde{F}_g$ and the renormalized ones in $\overline{\textrm{MS}}$ is given by:
\begin{eqnarray}
\tilde{F}^q(x) &=& F^q(x) 
- \left(\frac{1}{\epsilon} +\frac{1}{2} \ln\frac{e^\gamma \mu_F^2}{4\pi\mu_R^2} \right) K^{qq}(x,x')\otimes F^q(x') 
\nonumber \\ 
& &\phantom{F^q(x)}
- \left(\frac{1}{\epsilon} +\frac{1}{2} \ln\frac{e^\gamma \mu_F^2}{4\pi\mu_R^2} \right) K^{qg}(x,x')\otimes F^g(x')\,, \nonumber \\
%%%%%%%%%%%%%%%%%%%%%%%%%%%%%%%%%%%%%%%%%%%%%%%%%%%%%%%%%%%%
\tilde F^g(x) &=& F^g(x) 
- \left(\frac{1}{\epsilon} +\frac{1}{2} \ln\frac{e^\gamma \mu_F^2}{4\pi\mu_R^2} \right) K^{gg}(x,x')\otimes F^g(x')
\nonumber \\ 
& &\phantom{F^g(x)}
- \left(\frac{1}{\epsilon} +\frac{1}{2} \ln\frac{e^\gamma \mu_F^2}{4\pi\mu_R^2} \right) K^{gq}(x,x')\otimes F^q(x') \,,
\label{eq:evolution}
\end{eqnarray}
where evolution kernels $K^{qq},K^{qg}, K^{gg}, K^{gq}$ may be read from \cite{gpdrev}, $\otimes$ stands for integration over common variable from -1 to 1. At the NLO of the process studied in this paper the parts with $K^{gg}$ and $K^{gq}$ do not contribute, since the gluon contribution is of the $\mathcal{O}(\alpha_S)$. 
%%%%%%%%%%%%%%%%%%%%%%%%%%%%%%%%%%%%%%%%%%%%%%%%%%%%%%%%%%%%%%%%%%%%%%%%%%

In Eq. (\ref{eq:factorization}) unrenormalized coefficient functions contain infrared divergencies, and are given by:
\begin{eqnarray}
\tilde{T}^q &=&
C_0^q + 
\left(
\frac{\left|Q^2\right|e^\gamma}{4\pi\mu_R^2}
\right)^{\epsilon/2}
\left(
\frac{1}{\epsilon}~C_{coll}^{q} + C_1^{q}
\right)\,,
\nonumber\\
\tilde{T}^g&=&
\left(
\frac{\left|Q^2\right|e^\gamma}{4\pi\mu_R^2}
\right)^{\epsilon/2}
\left(
\frac{1}{\epsilon}~C_{coll}^{g} + C_1^{g}
\right)\,.
\label{eq:coeff}
\end{eqnarray}
$\tilde{T}^q$ is calculated using the following relation with $q\gamma \to q \gamma$ hard scattering amplitude $\mathcal{M^{\mu\nu}}$, given by diagrams shown on Figs. \ref{fig:selfenergy}, \ref{Fig:RV} and \ref{qbox} without attachement of external spinors of the t-channel quarks
%%%%%%%%%%%%%%%%%%%%%%%%%%%%%%%%%%%%%%%%%%%%%
\begin{eqnarray}
\tilde{T}^q &=&2\frac{g_T^{\mu\nu}}{D-2}\tr\left[\mathcal{M}_{\mu\nu}\frac{\not\!{p}}{4}\right]\,.
\label{eq:Hardscattampq}
\end{eqnarray}
In (\ref{eq:Hardscattampq}) the factor $\frac{g_T^{\mu\nu}}{D-2}$ is a part of the projection operator in Lorentz indices  on the symmetric part of the  full Compton scattering amplitude in Eq. (\ref{eq:factorization}). Factor $2$ is related to the definition of quark $F^q$ given by formula (\ref{eq:GPD}).

$\tilde{T}^g$ is calculated using the following relation with $g\gamma \to g \gamma$ hard scattering amplitude $\mathcal{M^{\mu\nu\alpha\beta}}$, given by diagrams shown on Figs. \ref{fig:Gluona} and \ref{fig:gluonb} without attachement of external polarization vectors of the t-channel gluons
\begin{eqnarray}
\tilde{T}^g&=&\frac{1}{2}\frac{-2x}{(x-x_B +i\varepsilon)(x+x_B-i\varepsilon)}\frac{g_T^{\mu\nu}}{(D-2)} \mathcal{M}_{\mu\nu\alpha\beta}\frac{g^{\alpha\beta}_T}{(D-2)} \,.
\label{eq:Hardscattampg}
\end{eqnarray}
Similarly to the quark case, factors $\frac{g_T^{\mu\nu}}{D-2}$  and $\frac{g^{\alpha\beta}_T}{(D-2)}$ are parts of the projector operators on symmetric two photon and two gluon states, respectively. The factor $\frac{1}{2}$ is the combinatorial factor which appears due to the condition that on the gluonic target we reproduce usual contribution of six diagrams shown in Figs. \ref{fig:Gluona} and \ref{fig:gluonb}. The factor $\frac{-2x}{(x-x_B +i\varepsilon)(x+x_B-i\varepsilon)}$ requires more explanations. It appears since in the axial gauge $n\cdot A =0$ we have the relation:
\begin{eqnarray}
\hspace*{-0.8cm}
&&\bigg\langle P' \bigg| 
A_a^{\alpha}\left(\frac{\lambda}{2}n\right)A_{a}^{\beta}\left(-\frac{\lambda}{2}n\right)
\bigg| P ~\bigg\rangle  {g_{T}}_{\alpha \beta}
\nonumber \\
&&
=\frac{-2x}{(x-x_B +i\varepsilon)(x+x_B-i\varepsilon)}
\bigg\langle P' \bigg|  F^{\mu \alpha}_a  \left( \frac{\lambda}{2}n \right)  F^{\nu\beta}_{a } \left( - \frac{\lambda}{2} n\right)\bigg| P ~\bigg\rangle  n_\mu n_\nu {g_{T}}_{\alpha \beta} \,.
\label{eq:tensortofield}
\end{eqnarray}
The structure of denominators in (\ref{eq:tensortofield}) is not fixed by the gauge condition $n \cdot A =0$ alone. This arbitrariness is due to the presence of the residual gauge. It is fixed by additional boundary conditions involved in the factorization procedure of the whole scattering amplitude of the given process. Here we fix it in agreement with the structure of denominators in the quark Born coefficient function for general double DVCS kinematics:
\begin{eqnarray}
C_{0~(DDVCS)}^q = e_q^2\left(\frac{1}{x-x_B+i\epsilon}+\frac{1}{x+x_B-i\epsilon}\right) \,.
\label{eq:BornCF}
\end{eqnarray}
In particular, in the case of DVCS where $x_B =\xi$ we obtain the standard expression 
$(x-\xi +i\varepsilon)(x+\xi-i\varepsilon)$ (see \cite{gpdrev}). In the case of the TCS where $x_B = -\xi$ this product becomes $(x+\xi +i\varepsilon)(x-\xi-i\varepsilon)$. Detailed calculation of $\tilde{T}^q$ and $\tilde{T}^g$ will be presented in the section \ref{sec:integrals}. 

%%%%%%%%%%%%%%%%%%%%%%%%%%%%%%%%%%%%%%%%%%%%%%%%%%%%
If the following relations between Born coefficient function, infrared divergent terms and evolution kernels hold:
\begin{eqnarray}
C_{coll}^{q}(x') &=& C_0^q(x) \otimes K^{qq}(x,x') \,,\nonumber\\
C_{coll}^{g}(x') &=& C_0^q(x) \otimes K^{qg}(x,x') \,.
\label{eq:factcond}
\end{eqnarray}
one can rewrite the full amplitude in the fully factorized form:
\begin{eqnarray}
\mathcal{A}^{\mu\nu} = g_T^{\mu\nu}\int_{-1}^1 dx 
\left[
\sum_q^{n_F} T^q(x) F^q(x)+T^g(x) F^g(x)
\right]\,,
\label{eq:factorizedamplitude}
\end{eqnarray}
where renormalized coefficient functions are given by:
\begin{eqnarray}
T^q&=& C_{0}^q +C_1^q +\frac{1}{2}\ln\left(\frac{|Q^2|}{\mu^2_F}\right) \cdot C_{coll}^q \,,\nonumber\\
T^g &=&  C_1^g +\frac{1}{2}\ln\left(\frac{|Q^2|}{\mu^2_F}\right) \cdot C_{coll}^g\,.
\label{eq:ceofficients}
\end{eqnarray} 
In the next section we will describe one-loop calculations necessary to obtain the above coefficient functions, in more details, as they can be useful in the calculations of similar processes (for example \cite{BDPpi}).
\section{Integrals}
\label{sec:integrals}
\subsection{Integrals with two propagators.}
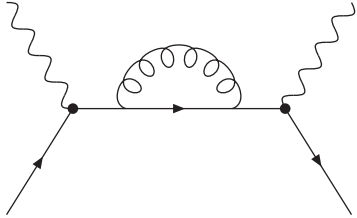
\begin{figure}[ht]
\begin{center}
\begin{picture}(150,100)(0,0)
%%%%%%%%%%%%%%1%%%%%%%%%%%%%%%%%%%%%%
\Vertex(35,50){2}
\Photon(10,90)(35,50){3}{4}
\ArrowLine(10,10)(35,50)
\ArrowLine(35,50)(115,50)
\Vertex(115,50){2}
\Photon(140,90)(115,50){3}{4}
\ArrowLine(115,50)(140,10)
\GlueArc(75,50)(20,0,180){4}{6}
\end{picture}
\end{center}
\caption{Self energy correction to $q\gamma \to q \gamma$ scattering amplitude}
\label{fig:selfenergy}
\end{figure}

We start with a detailed description of the diagram shown on Fig. \ref{fig:selfenergy}. Although this calculation is very simple, it reveals some characteristic features of the full calculation, and some pattern of the analytical structure of the result.

The symmetric part of the amplitude is given by:
\begin{eqnarray}
\tr\left[\mathcal{M}^{\mu\nu}\not\!{p}\right]&=& ie^2g^2C_F\frac{1}{[(q+xp)^2+i\varepsilon]^2}
\int(dk) \frac{\tr[\gamma^\mu(\not\!{q} + x\not\!{p}) \gamma^\rho(\not\!{q} + x\not\!{p}+\not\!{k})
\gamma_\rho(\not\!{q} + x\not\!{p})\gamma^\nu\not\!p]}{[(k+q+xp)^2 +i\varepsilon][k^2+i\varepsilon]} \,,
\label{eq:SE1_symm}
\end{eqnarray}
where $(dk) \equiv \mu^{4-D}\frac{d^Dk}{(2\pi)^D}$, and $C_F= \frac{N^2-1}{2N}$. We have two types of integrals to perform:
\begin{eqnarray}
b_0 &\equiv & \int(dk)\frac{1}{[(k+q+xp)^2 +i\varepsilon][k^2+i\varepsilon]}\,,\nonumber\\
b_\sigma&\equiv & \int(dk)\frac{k_\sigma}{[(k+q+xp)^2 +i\varepsilon][k^2+i\varepsilon]} = -\frac{1}{2}(q+xp)_\sigma b_0 \,,
\label{eq:b0def}
\end{eqnarray}
$b_0$ may be shown to be equal (pay attention to the difference between $\epsilon$ and $\varepsilon$):
\begin{eqnarray}
b_0 &=&  \frac{i}{(4\pi)^2} \frac{1}{(4 \pi\mu^2)^{\frac{\epsilon}{2}}}\Gamma\left(-\frac{\epsilon}{2}\right)
\frac{\Gamma\left(1+\frac{\epsilon}{2}\right)\Gamma\left(1+\frac{\epsilon}{2}\right)}{\Gamma\left(2+\epsilon\right)}
\left(Q^2\frac{x_B-x}{x_B}-i\varepsilon\right)^{\frac{\epsilon}{2}}\,.
\label{eq:b0}
\end{eqnarray}
When we add the diagram with external photon lines crossed, which is given by the $x_B \leftrightarrow -x_B$ substitution, we get the following result for the sum of those two diagrams with the self-energy corrections:
\begin{eqnarray}
\tr\left[\mathcal{M}_\Sigma^{\mu\nu}\not\!{p}\right]
&=&- g_T^{\mu\nu}\frac{e^2\alpha_s C_F}{4\pi}\frac{1}{(4 \pi \mu^2)^{\frac{\epsilon}{2}}}
\bigg\{\phantom{-}~ \frac{Q^2}{x_B}\frac{1}{[Q^2\frac{x-x_B}{x_B}+i\varepsilon]} 
\left(-\frac{4}{\epsilon}+2\right)\left(Q^2\frac{x_B-x}{x_B}-i\varepsilon\right)^{\frac{\epsilon}{2}}
\nonumber\\
& &\phantom{- g_T^{\mu\nu}\frac{e^2\alpha_s C_F}{4\pi}\frac{1}{(4 \pi \mu^2)^{\frac{\epsilon}{2}}}
\bigg\{}
 - \frac{Q^2}{x_B}\frac{1}{[-Q^2\frac{x+x_B}{x_B}+i\varepsilon]} 
\left(-\frac{4}{\epsilon}+2\right)\left(Q^2\frac{x_B+x}{x_B}-i\varepsilon\right)^{\frac{\epsilon}{2}}
\bigg\}\,,
\label{eq:SEgen}
\end{eqnarray}
which in the DVCS and TCS limits results in:
\begin{eqnarray}
\tr\left[\mathcal{M}_\Sigma^{\mu\nu}\not\!{p}\right]_{DVCS}
&=&
- g_T^{\mu\nu}\frac{e^2\alpha_s C_F}{4\pi}\left(\frac{Q^2}{4 \pi \mu^2}\right)^{\frac{\epsilon}{2}}2~\bigg\{
\phantom{+}~\frac{1}{[x+\xi-i\varepsilon]} 
\left[-\frac{2}{\epsilon}+1-\log\left(1 +\frac{x}{\xi}-i\varepsilon\right)\right]
\nonumber\\
& &
\phantom{
- g_T^{\mu\nu}\frac{e^2\alpha_s C_F}{4\pi}\left(\frac{Q^2}{4 \pi \mu^2}\right)^{\frac{\epsilon}{2}}2~\bigg\{}
+\frac{1}{[x-\xi+i\varepsilon]} 
\left[-\frac{2}{\epsilon}+1 - \log\left(1 -\frac{x}{\xi}-i\varepsilon\right)\right]
\bigg\}\,,\nonumber\\
%\label{eq:com}
%\end{eqnarray}
%\begin{eqnarray}
\tr\left[\mathcal{M}_\Sigma^{\mu\nu}\not\!{p}\right]_{TCS}
&=&- g_T^{\mu\nu}\frac{e^2\alpha_s C_F}{4\pi}\left(\frac{Q'^2}{4 \pi \mu^2}\right)^{\frac{\epsilon}{2}}
2~\bigg\{\phantom{-}~ \frac{1}{[x-\xi-i\varepsilon]} 
\left[-\frac{2}{\epsilon}+1-\log\left(-1 +\frac{x}{\xi}-i\varepsilon\right)\right]
\nonumber\\
& &
\phantom{- g_T^{\mu\nu}\frac{e^2\alpha_s C_F}{4\pi}\left(\frac{Q'^2}{4 \pi \mu^2}\right)^{\frac{\epsilon}{2}}
2~\bigg\{}
+ \frac{1}{[x+\xi+i\varepsilon]} 
\left[-\frac{2}{\epsilon}+1 - \log\left(-1 -\frac{x}{\xi}-i\varepsilon\right)\right]
\bigg\} \,.\nonumber\\
\label{eq:SElimits}
\end{eqnarray}
We notice that in the TCS case we have $\xi + i\varepsilon$ contrarily to the $\xi - i\varepsilon$ present in the DVCS. There is also an overall minus sign under the logarithm, coming from the different sign of $Q^2$.
%%%%%%%%%%%%%%%%%%%%%%%%%%%%%%%%%%%%%%%%%%%%%%%%%%%%%%%%%%%%%%%%%%%%%%%%%%%%%%%%%%%%%%%%%%%%%%%%%%%%%%%%%%%%%
%%%%%%%%%%%%%%%%%%%%%%%%%%%%%%%%%%%%%%%%%%%%%%%%%%%%%%%%%%%%%%%%%%%%%%%%%%%%%%%%%%%%%%%%%%%%%%%%%%%%%%%%%%%%%
\subsection{Integrals with three propagators.}
\begin{figure}[ht]
\begin{center}
\begin{picture}(150,100)(0,0)
%%%%%%%%%%%%%%1%%%%%%%%%%%%%%%%%%%%%%
\Vertex(35,50){2}
\Photon(10,90)(35,50){3}{4}
\ArrowLine(10,10)(35,50)
\ArrowLine(35,50)(115,50)
\Vertex(115,50){2}
\Photon(140,90)(115,50){3}{4}
\ArrowLine(115,50)(140,10)
\GlueArc(110,50)(20,180,310){4}{6}
\end{picture}
\end{center}
\caption{Right vertex correction to $q\gamma \to q \gamma$ scattering amplitude}
\label{Fig:RV}
\end{figure}
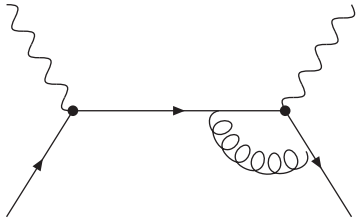

In this section we will describe in a detailed way the calculation of the diagram shown on Fig. \ref{Fig:RV}, because all other diagrams with three, and some of the diagrams with four propagators in the loop, may be calculated in a similar way.
The symmetric part of the amplitude with the right vertex correction is given by:
\begin{eqnarray}
\tr\left[\mathcal{M}^{\mu\nu}_{RV} \not \! p\right] &=& ie^2 g^2 C_F \frac{1}{(q+xp)^2+i\varepsilon}
\int (dk)
\frac{
\tr\left[\gamma^\rho 
(\not\! k - \xi \not \! p) \gamma^\nu (\not\! k +\not \! q ) \gamma_\rho(\not\! q + x \not \! p)\gamma^{\mu} \not\!p \right]
}
{[(k+q)^2 + i\varepsilon][(k-\xi p)^2 + i\varepsilon][(k-xp)^2 + i\varepsilon]} \,.\nonumber \\
\label{eq:RV}
\end{eqnarray}
We start with the integration over $k^-$. There are three poles placed at:
\begin{eqnarray}
k_1^- &=&\frac{k_\bot^2 -i \varepsilon}{2(y-x)} \,,
%\quad\textrm{ in the lower halfplane for } y>x 
\nonumber \\
k_2^- &=&\frac{k_\bot^2 -i \varepsilon}{2(y-\xi)} \,,
%\quad \textrm{ in the lower halfplane for } y>\xi 
\nonumber \\
k_3^- &=&\frac{k_\bot^2 +Q^2 \left(1-\frac{y}{x_B}\right)-i \varepsilon}{2(y-x_B)} \,.
%\quad\textrm{ in the lower halfplane for } y>x_B\nonumber
\label{eq:poles}
\end{eqnarray}
For various values of $y$ we close the contours of integration in the upper or lower half plane, in such a way that we avoid catching the $k_3^-$ pole. Irrespectively of the ordering of $x$, $x_B$ and $\xi$, we arrive at:
\begin{eqnarray} 
{g_T}_{\mu\nu}\tr\left[\mathcal{M}^{\mu\nu}_{RV} \not \! p\right]  &=& ie^2 g^2 C_F \frac{1}{(q+xp)^2+i\varepsilon}
\left(
-i \int_x^{x_B} \frac{dy}{2\pi}\int (d k_\bot) \textrm{~Res}_{k_1^-} f
-i \int_\xi^{x_B} \frac{dy}{2\pi}\int (d k_\bot) \textrm{~Res}_{k_2^-} f
\right)\,,
\nonumber \\
\end{eqnarray}
where $(d k_\bot) = \mu^{-\epsilon} \frac{d^{D-2}k_\bot}{(2\pi)^{D-2}}$ and $y\equiv k^+$. Residua of the first and the second pole are given by:
\begin{eqnarray}
\textrm{~Res}_{k_1^-} f &=& \frac{y-x}{2(x-x_B)(x-\xi)} \cdot
\frac{\alpha_1 + \beta_1 k_\bot^2}
{\left[k_\bot^2 \right]
\left[k_\bot^2 - Q^2 \frac{(x_B-y)(y-x)}{x_B(x-x_B)}-i \varepsilon\right]
}\,,
%\label{eq:Residuum1}
%\end{eqnarray}
%\begin{eqnarray}
\nonumber \\
\textrm{~Res}_{k_2^-} f &=& \frac{y-\xi}{2(\xi-x_B)(\xi-x)} \cdot
\frac{\alpha_2 + \beta_2 k_\bot^2}
{\left[k_\bot^2 \right]
\left[k_\bot^2 - Q^2 \frac{(x_B-y)(y-\xi)}{x_B(\xi-x_B)}-i \varepsilon\right]
}\,,
\label{eq:ResiduaRV}
\end{eqnarray}
where $\alpha_i$ and $\beta_i$ are defined by the value of the numerator at the correspondent pole:
\begin{eqnarray}
\alpha_i + \beta_i k_\bot^2 &\equiv&
 \tr\left[\gamma^\rho 
(\not\! k - \xi \not \! p) \gamma^\nu (\not\! k +\not \! q ) \gamma_\rho(\not\! q + x \not \! p)\gamma^{\mu} \not\!p \right] \Bigg|_{k^-_i} \,.
\label{eq:alphabeta}
\end{eqnarray}
After we perform the integration over $k^-$ we arrive to:
\begin{eqnarray}
{g_T}_{\mu\nu}\tr\left[\mathcal{M}^{\mu\nu}_{RV} \not \! p\right]
&=& e^2 g^2 C_F \frac{1}{(q+xp)^2+i\varepsilon}
\nonumber \\
& &
\bigg(\int_x^{x_B} \frac{dy}{2\pi}\int (d k_\bot)
\frac{y-x}{2(x-x_B)(x-\xi)} \cdot
\frac{\alpha_1 + \beta_1 k_\bot^2}
{\left[k_\bot^2 \right]
\left[k_\bot^2 - Q^2 \frac{(x_B-y)(y-x)}{x_B(x-x_B)}-i \varepsilon\right]
}
\nonumber\\ 
&+&\int_\xi^{x_B} \frac{dy}{2\pi}\int (d k_\bot) \frac{y-\xi}{2(\xi-x_B)(\xi-x)} \cdot
\frac{\alpha_2 + \beta_2 k_\bot^2}
{\left[k_\bot^2 \right]
\left[k_\bot^2 - Q^2 \frac{(x_B-y)(y-\xi)}{x_B(\xi-x_B)}-i \varepsilon\right]
}
\bigg) \,.
\label{eq:intkty}
\end{eqnarray}
All integrals in $k_\bot$ we encounter during the calculation have the following form: 
\begin{eqnarray}
\int\frac{d^{D-2}k_\bot}{(2\pi)^{D-2}}
\frac{\alpha + \beta k_\bot^2 }
{\left[k_\bot^2 - i \varepsilon\right]
\left[k_\bot^2 - M^2-i \varepsilon\right]}  
&=& \left( \frac{\alpha+ \beta M^2}{M^2}\right)  a(M^2)\,,
\nonumber\\
\int\frac{d^{D-2}k_\bot}{(2\pi)^{D-2}}
\frac{\alpha + \beta k_\bot^2 }
{\left[k_\bot^2 -M_1^2-i \varepsilon\right]
\left[k_\bot^2 - M_2^2-i \varepsilon\right]}  
&=& \left( \frac{\alpha+ \beta M_1^2}{M_1^2-M_2^2}\right)  a(M_1^2)
   -\left( \frac{\alpha+ \beta M_2^2}{M_1^2-M_2^2}\right)  a(M_2^2) \,,
\nonumber\\
a(M^2)&=&
\frac{1}{(4\pi)^{\frac{D-2}{2}}}\left(-M^2-i\varepsilon \right)^{\frac{D-4}{2}} \Gamma \left(\frac{4-D}{2}\right)\,, \nonumber\\
%&+& \gamma \frac{1}{(4\pi)^{\frac{D-2}{2}}}\left(-M^2-i\varepsilon \right)^{\frac{D-2}{2}} 
%\frac{\Gamma \left(\frac{D}{2}\right)\Gamma \left(\frac{2-D}{2}\right)}{\Gamma \left(\frac{D-2}{2}\right) }  
\label{eq:ktinegrb}
\end{eqnarray}
so after $k_\bot$ integration we get the following result:
\begin{eqnarray}
{g_T}_{\mu\nu}\tr\left[\mathcal{M}^{\mu\nu}_{RV} \not \! p\right]  
&=&
\frac{e^2 \alpha_S C_F}{4\pi}\left(\frac{1}{4\pi\mu^2}\right)^{\frac{\epsilon}{2}} 
\frac{1}{(q+xp)^2+i\varepsilon}\Gamma\left(-\frac{\epsilon}{2}\right) \frac{1}{x-\xi}\cdot
\nonumber \\
& &
\bigg[-\left(Q^2\frac{x_B-x}{x_B}-i\varepsilon\right)^{\frac{\epsilon}{2}} 
\int_x^{x_B} dy
\left(\frac{y-x}{x_B-x}\right)^{1+\frac{\epsilon}{2}}  \cdot
\left(\frac{x_B-y}{x_B-x}\right)^{\frac{\epsilon}{2}} 
\left(\frac{\alpha_1}{M_1^2} + \beta_1\right)
\nonumber\\ 
&+&
\left(Q^2\frac{x_B-\xi}{x_B}-i\varepsilon\right)^{\frac{\epsilon}{2}} 
\int_\xi^{x_B} dy
\left(\frac{y-\xi}{x_B-\xi}\right)^{1+\frac{\epsilon}{2}}  \cdot
\left(\frac{x_B-y}{x_B-\xi}\right)^{\frac{\epsilon}{2}} 
\left(\frac{\alpha_2}{M_2^2} + \beta_2\right)
\bigg]  \,.\nonumber \\
\label{eq:inty}
\end{eqnarray}
The last integration is performed making use of the beta function definition. The diagram with a left vertex correction is given by symmetry $\xi \to -\xi$, and the crossed diagrams by $x_B \to -x_B$. When we include all four vertex corrections:
\begin{eqnarray}
\tr\left[\mathcal{M}^{\mu\nu}_{V} \not \! p\right]  &=& 
\left(\tr\left[\mathcal{M}^{\alpha\beta}_{RV} \not \! p\right]  + (\xi \to -\xi)\right)+  (x_B \to -x_B)\,,
\end{eqnarray}
we get a result with the following structure:
\begin{eqnarray}
\tr\left[\mathcal{M}^{\mu\nu}_{V} \not \! p\right]  &=& g_T^{\mu\nu}\frac{e^2 \alpha_S C_F}{4\pi}\left(\frac{1}{4\pi\mu^2}\right)^{\frac{\epsilon}{2}}
\bigg\{ \nonumber\\
& &\left(Q^2\frac{x_B-x}{x_B}-i\varepsilon\right)^{\frac{\epsilon}{2}} 
\frac{1}{Q^2\frac{x-x_B}{x_B}+i\varepsilon}\bigg[f_1(x_B,\xi,x,\epsilon,\varepsilon)+f_1(x_B,-\xi,x,\epsilon,\varepsilon) \bigg]
\nonumber\\
%&+&\left(Q^2\frac{x_B+x}{x_B}-i\varepsilon\right)^{\frac{\epsilon}{2}} 
%\frac{1}{-Q^2\frac{x+x_B}{x_B}+i\varepsilon}\bigg[f_1(-x_B,\xi,x,\epsilon)+f_1(-x_B,-\xi,x,\epsilon) \bigg]
%\nonumber\\
&+&\left(Q^2\frac{x_B-\xi}{x_B}-i\varepsilon\right)^{\frac{\epsilon}{2}} 
\bigg[\frac{1}{Q^2\frac{x-x_B}{x_B}+i\varepsilon}f_2(x_B,\xi,x,\epsilon,\varepsilon)
+\frac{1}{-Q^2\frac{x+x_B}{x_B}+i\varepsilon}f_2(-x_B,-\xi,x,\epsilon,\varepsilon)\bigg]
\nonumber\\
%&+&\left(Q^2\frac{x_B+\xi}{x_B}-i\varepsilon\right)^{\frac{\epsilon}{2}} 
%\bigg[\frac{1}{Q^2\frac{x-x_B}{x_B}+i\varepsilon}f_2(x_B,-\xi,x,\epsilon)
%+\frac{1}{-Q^2\frac{x+x_B}{x_B}+i\varepsilon}f_2(-x_B,\xi,x,\epsilon)\bigg] 
&& +( x_B \leftrightarrow -x_B)\bigg\}\,,
\label{eq:Vertexstructure}
\end{eqnarray}
where $f_1$ and $f_2$ are some complicated functions of $x_B$, $\xi$, $x$, $\epsilon$, $\varepsilon$.
\subsection{Integrals with four propagators.}
All of the integrals with four propagators may be reduced to the three propagator case, although some of them require some care.  
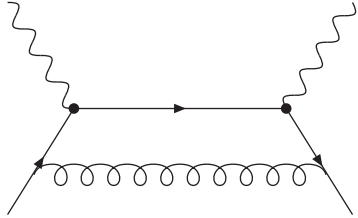
\begin{figure}[ht]
\begin{center}
\begin{picture}(150,100)(0,0)
%%%%%%%%%%%%%%1%%%%%%%%%%%%%%%%%%%%%%
\Vertex(35,50){2}
\Photon(10,90)(35,50){3}{4}
\ArrowLine(10,10)(35,50)
\ArrowLine(35,50)(115,50)
\Vertex(115,50){2}
\Photon(140,90)(115,50){3}{4}
\ArrowLine(115,50)(140,10)
\Gluon(20,25)(130,25){4}{10}
\end{picture}
\caption{\label{qbox}Box diagram correction to $q\gamma \to q \gamma$ scaterring amplitude}
\end{center}
\end{figure}
We will start this section with the calculation of the symmetric part of the box diagram shown in the Fig. \ref{qbox}:
\begin{eqnarray}
\tr \left[\mathcal{M}^{\mu\nu}_B \not\! p\right] = ig^2e^2C_F\int (dk)
\frac{\tr\left[\gamma^\rho(\not\!k -\xi \not\!p)\gamma^\nu(\not\!k + \not\!q)
\gamma^\mu(\not\!k +\xi \not\!p)\gamma_\rho\not\!p\right]}
{[(k-\xi p)^2 + i \varepsilon][(k+\xi p)^2 + i \varepsilon][(k+q)^2 + i \varepsilon][(k-x p)^2 + i \varepsilon]} \,.
\label{eq:boxtr}
\end{eqnarray}
In this case we have four denominators in the integrated function, but one can easily check that: 
\begin{eqnarray}
{g_T}_{\mu\nu}\tr\left[\gamma^\rho(\not\!k -\xi \not\!p)\gamma^\nu(\not\!k + \not\!q)
\gamma^\mu(\not\!k +\xi \not\!p)\gamma_\rho\not\!p\right]\equiv A\cdot k^2+B \cdot 2kp \,,
\label{eq:dec}
\end{eqnarray}
so, using following relations:
\begin{eqnarray}
k^2 &=& \frac{1}{2}(k+\xi p )^2 + (\xi \to -\xi) \,,\nonumber \\
2 k\cdot p &=& \frac{1}{2\xi}(k+\xi p )^2 + (\xi \to -\xi)\,,
\end{eqnarray}
one can easily decompose the four denominator integral into two integrals with three denominators:
\begin{eqnarray}
{g_T}_{\mu\nu}\tr \left[\mathcal{M}^{\mu\nu}_B \not\! p\right] 
&=& ig^2e^2C_F\frac{1}{2}\int (dk)\frac{A+\frac{1}{\xi}B}
{\left[(k-xp)^2+ i \varepsilon\right]
\left[(k-\xi p)^2+ i \varepsilon\right]
\left[(k + q)^2+ i \varepsilon\right]}+ (\xi \to -\xi) \,,\nonumber
\label{eq:reduction}
\end{eqnarray}
which we calculate in the same way as the vertex corrections diagrams. Crossed diagram is given by the $x_B$ to $-x_B$ replacement.
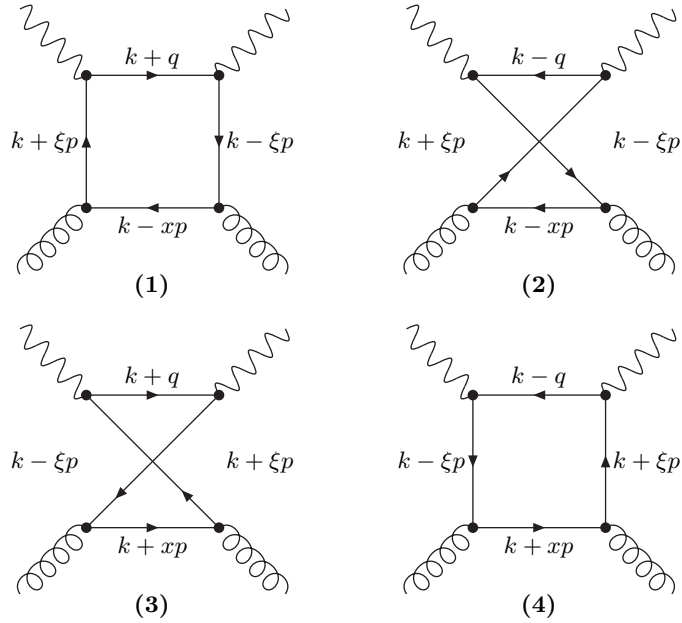
\begin{figure}[h!]
\begin{center} 
\begin{picture}(100,100)(0,0)
\ArrowLine(25,25)(25,75)
\ArrowLine(25,75)(75,75)
\ArrowLine(75,75)(75,25)
\ArrowLine(75,25)(25,25)
\Gluon(0,0)(25,25){4}{4}  \Vertex(25,25){2}
\Gluon(75,25)(100,0){4}{4}  \Vertex(75,25){2}
\Photon(0,100)(25,75){4}{4} \Vertex(25,75){2}
\Photon(75,75)(100,100){4}{4} \Vertex(75,75){2}
\Text(50,22)[t]{$k-xp$}
\Text(22,50)[r]{$k+\xi p$}
\Text(78,50)[l]{$k-\xi p$}
\Text(50,78)[b]{$k+ q $}
\Text(50,0)[t]{\bf(1)}
\end{picture}
%%%%%%%%%%%%%%%%%%%%%%%%%%%%%%%%%%%%%%%     2    %%%%%%%%%%%%%%%%%%%%%%%%%%%%%%%%%%%%%%%%%%%%%%%%%
\hspace{40 px}
\begin{picture}(100,100)(0,0)
\ArrowLine(25,25)(50,50)
\Line(50,50)(75,75)
\ArrowLine(75,75)(25,75)
\ArrowLine(50,50)(75,25)
\Line(25,75)(50,50)
\ArrowLine(75,25)(25,25)
\Gluon(0,0)(25,25){4}{4}  \Vertex(25,25){2}
\Gluon(75,25)(100,0){4}{4}  \Vertex(75,25){2}
\Photon(0,100)(25,75){4}{4} \Vertex(25,75){2}
\Photon(75,75)(100,100){4}{4} \Vertex(75,75){2}
\Text(50,22)[t]{$k-xp$}
\Text(22,50)[r]{$k+\xi p$}
\Text(78,50)[l]{$k-\xi p$}
\Text(50,78)[b]{$k - q $}
\Text(50,0)[t]{\bf(2)}
\end{picture}

\vspace{20 px}
%%%%%%%%%%%%%%%%%%%%%%%%%%%%%%%%%%%%%%%     3    %%%%%%%%%%%%%%%%%%%%%%%%%%%%%%%%%%%%%%%%%%%%%%%%%
\begin{picture}(100,100)(0,0)
\ArrowLine(50,50)(25,25)
\Line(50,50)(75,75)
\ArrowLine(25,75)(75,75)
\ArrowLine(75,25)(50,50)
\Line(25,75)(50,50)
\ArrowLine(25,25)(75,25)
\Gluon(0,0)(25,25){4}{4}  \Vertex(25,25){2}
\Gluon(75,25)(100,0){4}{4}  \Vertex(75,25){2}
\Photon(0,100)(25,75){4}{4} \Vertex(25,75){2}
\Photon(75,75)(100,100){4}{4} \Vertex(75,75){2}
\Text(50,22)[t]{$k+xp$}
\Text(22,50)[r]{$k-\xi p$}
\Text(78,50)[l]{$k+\xi p$}
\Text(50,78)[b]{$k+ q $}
\Text(50,0)[t]{\bf(3)}
\end{picture}
%%%%%%%%%%%%%%%%%%%%%%%%%%%%%%%%%%%%%%%     4    %%%%%%%%%%%%%%%%%%%%%%%%%%%%%%%%%%%%%%%%%%%%%%%%%
\hspace{40 px}
\begin{picture}(100,100)(0,0)
\ArrowLine(25,75)(25,25)
\ArrowLine(75,75)(25,75)
\ArrowLine(75,25)(75,75)
\ArrowLine(25,25)(75,25)
\Gluon(0,0)(25,25){4}{4}  \Vertex(25,25){2}
\Gluon(75,25)(100,0){4}{4}  \Vertex(75,25){2}
\Photon(0,100)(25,75){4}{4} \Vertex(25,75){2}
\Photon(75,75)(100,100){4}{4} \Vertex(75,75){2}
\Text(50,22)[t]{$k+xp$}
\Text(22,50)[r]{$k-\xi p$}
\Text(78,50)[l]{$k+\xi p$}
\Text(50,78)[b]{$k- q $}
\Text(50,0)[t]{\bf(4)}
\end{picture}
\caption{First group of diagrams describing $\gamma g \to \gamma g$ scattering.}
\label{fig:Gluona}
\vspace{20 px}
\end{center}
\end{figure}

Let us now turn to the gluon coefficient functions. The symmetric part of the first diagram describing $\gamma g \to \gamma g$  scattering, shown on the Fig. \ref{fig:Gluona}, is given by:
\begin{eqnarray}
g_T^{\mu\nu}g_T^{\alpha\beta}  \mathcal{M}_{(1) \mu\nu\alpha\beta} 
=   i e^2g^2T_F \int (dk) \frac{
{g_T}_{\alpha\beta}{g_T}_{\mu\nu}\cdot\tr\left[
\gamma^\alpha (\not \! k - x\not \!p)
\gamma^\beta  (\not \! k - \xi\not \!p)
\gamma^\mu (\not \! k +\not \!q)
\gamma^\nu (\not \! k +\xi \not \!p)
\right]}
{\left[(k-xp)^2+ i \varepsilon\right]
\left[(k-\xi p)^2+ i \varepsilon\right]
\left[(k + q)^2+ i \varepsilon\right]
\left[(k+\xi p)^2+ i \varepsilon\right]} \,,
\label{eq:diag}
\end{eqnarray}
with $T_F=\frac{1}{2}$. The structure of the numerator is similar to the one given by equation (\ref{eq:dec}). So we can use the same decomposition as in the case of the quark box diagram. Diagrams (2), (3) and (4) from Fig. \ref{fig:Gluona} are 
connected to diagram (1) by simple symmetries. To get diagram (2) one has to change $x_B \leftrightarrow -x_B$, 
diagram (3) $x \leftrightarrow -x$, and to get diagram (4) one has to do both changes.
\begin{figure}[b]
\begin{center} 
\begin{picture}(100,100)(0,0)
\ArrowLine(25,25)(25,75)
\ArrowLine(25,75)(50,50)
\Line(50,50)(75,25)
\ArrowLine(75,25)(75,75)
\ArrowLine(50,50)(25,25)
\Line(75,75)(50,50)
\Gluon(0,0)(25,25){4}{4}  \Vertex(25,25){2}
\Gluon(75,25)(100,0){4}{4}  \Vertex(75,25){2}
\Photon(0,100)(25,75){4}{4} \Vertex(25,75){2}
\Photon(75,75)(100,100){4}{4} \Vertex(75,75){2}
\Text(50,22)[t]{$D$}
\Text(22,50)[r]{$C$}
\Text(78,50)[l]{$A$}
\Text(50,78)[b]{$B $}
\Text(50,0)[t]{\bf(5)}
\end{picture}
%%%%%%%%%%%%%%%%%%%%%%%%%%%%%%%%%%%%%%%     2    %%%%%%%%%%%%%%%%%%%%%%%%%%%%%%%%%%%%%%%%%%%%%%%%%
\hspace{40 px}
\begin{picture}(100,100)(0,0)
\ArrowLine(25,75)(25,25)
\ArrowLine(50,50)(25,75)
\Line(75,25)(50,50)
\ArrowLine(75,75)(75,25)
\ArrowLine(25,25)(50,50)
\Line(50,50)(75,75)
\Gluon(0,0)(25,25){4}{4}  \Vertex(25,25){2}
\Gluon(75,25)(100,0){4}{4}  \Vertex(75,25){2}
\Photon(0,100)(25,75){4}{4} \Vertex(25,75){2}
\Photon(75,75)(100,100){4}{4} \Vertex(75,75){2}
\Text(50,22)[t]{$C_q$}
\Text(22,50)[r]{$D_q$}
\Text(78,50)[l]{$B_q$}
\Text(50,78)[b]{$A_q $}
\Text(50,0)[t]{\bf(6)}
\end{picture}
\caption{Second group of diagrams describing $\gamma g \to \gamma g$ scattering.}
\label{fig:gluonb}
\vspace{20 px}
\end{center}
\end{figure}
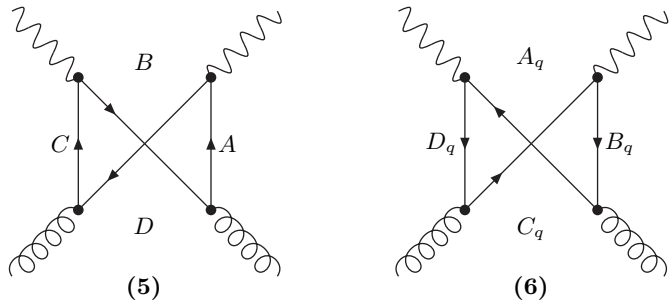

Diagrams shown on Fig. \ref{fig:gluonb}, have different denominator structure, so we will describe the way of dealing with them more precisly.  Momenta flowing in the diagram (5) may be chosen as:
\begin{eqnarray}
A&=&k+q-(x-\xi)p \,,\nonumber\\
B&=&k+q \,,\nonumber\\
C&=&k+\xi p \,,\nonumber \\
D&=&k-x p \,,
\label{eq:ABCD}
\end{eqnarray}
and $A_q, B_q, C_q, D_q$ from diagram (6) are equal to $A, B, C, D$ with $q \leftrightarrow -q$. Both diagrams give the same result:
\begin{eqnarray} 
g_T^{\mu\nu}g_T^{\alpha\beta}  \mathcal{M}_{(5) \mu\nu\alpha\beta} = ie^2g^2T_F \int (dk) 
\frac{g_{T \mu\nu}g_{T \alpha\beta}\tr \left[\gamma^\mu \not \! A \gamma^\beta \not \! B \gamma^\nu\not \! C \gamma^\alpha \not \! D \right]}
{[A^2+i \varepsilon][B^2+i \varepsilon][C^2+i \varepsilon][D^2+i \varepsilon]},. 
\label{eq:diag4b}
\end{eqnarray}
%where:
%\begin{eqnarray}
%NUM &=& g_{T \mu\nu}g_{T \alpha\beta}\tr \left[\gamma^\mu \not \! A \gamma^\beta \not \! B \gamma^\nu\not \! C \gamma^\alpha \not \! D \right]\nonumber\\
%DEN &=& [A^2+i \epsilon][B^2+i \epsilon][C^2+i \epsilon][D^2+i \epsilon]
%\label{eq:NUMDEN4b}
%\end{eqnarray}
As previously we notice that the numerator may be written as  $\mathcal{A}~k^2+\mathcal{B}~2k\cdot p \,.$
To reduce our integral to the three denominator case, we use other relations:
\begin{eqnarray}
k^2 &=& \frac{x}{x+\xi}(k + \xi p)^2+ \frac{\xi}{x+\xi}(k -x p)^2  \,,\nonumber \\
2 k \cdot p &=& \frac{1}{x+\xi}(k + \xi p)^2- \frac{1}{x+\xi}(k -x p)^2 \,, \nonumber
\label{eq:decomp}
\end{eqnarray}
so we end up with:
\begin{eqnarray}
I_{(5)} &=&   
\frac{1}{(x+\xi)}\int (dk) \frac{\mathcal{A}~x+\mathcal{B}}{[A^2+i \varepsilon][B^2+i \varepsilon][D^2+i \varepsilon]} 
+
\frac{1}{(x+\xi)}\int (dk) \frac{\mathcal{A}~\xi-\mathcal{B}}{[A^2+i \varepsilon][B^2+i \varepsilon][C^2+i \varepsilon]}
\nonumber\\
&=& I_1+I_2 \,.
\label{I1plusI2}
\end{eqnarray} 
One could worry if the above decomposition is well defined for $x = -\xi$, but it is easy to check, that the expression (\ref{I1plusI2}) is regular in that limit. 

As previously we start with integration over $k^-$. We find four poles:
\begin{eqnarray}
(k-xp)^2 +i\varepsilon =0 &\Rightarrow &k^-_1 
= \frac{k_\bot^2-i\varepsilon}{2(y-a_1)}  \,,\nonumber \\
(k+q)^2 +i\varepsilon =0 &\Rightarrow &k^-_2 
= \frac{k_\bot^2-i\varepsilon+Q^2\left(\frac{a_2-y}{x_B}\right)}{2(y-a_2)}  \,,\nonumber \\
(k+q-(x-\xi)p)^2 +i\varepsilon =0 &\Rightarrow &k^-_3 = \frac{k_\bot^2-i\varepsilon+Q^2\left(\frac{a_3-y}{x_B}\right)}{2(y-a_3)} \,, \nonumber \\
(k-\xi p)^2 +i\varepsilon =0 &\Rightarrow &k^-_4 = \frac{k_\bot^2-i\varepsilon}{2(y-a_4)}   \,,
\label{eq:fourpoles}
\end{eqnarray}
where $a_1= x$, $a_2= x_B$, $a_3= x_B+x-\xi$ and $a_4= -\xi$ are values of $y$ for which poles imaginary parts change sign. Again appropriate choice of the integration contours allows us to write:
\begin{eqnarray}
I_1 &=& -i\mu^{D-4} \int_{a_1}^{a_3}\frac{dy}{2\pi}  \int\frac{d^{D-2}k_\bot}{(2\pi)^{D-2}}  \textrm{~Res}_{k_1^-} f_1
%\nonumber \\& & 
-i\mu^{D-4} \int_{a_2}^{a_3}\frac{dy}{2\pi} \int\frac{d^{D-2}k_\bot}{(2\pi)^{D-2}} \textrm{~Res}_{k_2^-} f_1 \,, 
\nonumber \\
%\label{eq:I1}
%\end{eqnarray}
%\begin{eqnarray}
I_2 &=& -i\mu^{D-4} \int_{a_4}^{a_3}\frac{dy}{2\pi}  \int\frac{d^{D-2}k_\bot}{(2\pi)^{D-2}}  \textrm{~Res}_{k_4^-} f_2
%\nonumber \\& & 
-i\mu^{D-4} \int_{a_2}^{a_3}\frac{dy}{2\pi} \int\frac{d^{D-2}k_\bot}{(2\pi)^{D-2}} \textrm{~Res}_{k_2^-} f_2 
\,.
\label{eq:I1I2}
\end{eqnarray}
The only difference with equations (\ref{eq:ResiduaRV}) is that we now have additional mass term in the denominator:
\begin{eqnarray}
\textrm{~Res}_{k_1^-} f_1 &=& \frac{1}{x+\xi}\cdot\frac{y-a_1}{2(a_1-a_2)(a_1-a_3)} \cdot
\frac{\alpha_1 + \beta_1 k_\bot^2}
{\left[k_\bot^2 -M_{12}^2- i \varepsilon\right]
\left[k_\bot^2 - M_{13}^2-i \varepsilon\right]
}\,,
\nonumber\\
\textrm{~Res}_{k_2^-} f_1 &=& 
\frac{1}{x+\xi}\cdot\frac{y-a_2}{2(a_2-a_1)(a_2-a_3)} \cdot
\frac{\alpha_2 + \beta_2 k_\bot^2}
{\left[k_\bot^2 -M_{12}^2- i \varepsilon\right]
\left[k_\bot^2 -i \varepsilon\right]
}\,,
\nonumber\\
\textrm{~Res}_{k_2^-} f_2 &=& 
\frac{1}{x+\xi}\cdot\frac{y-a_2}{2(a_2-a_4)(a_2-a_3)} \cdot
\frac{\alpha_3 + \beta_3 k_\bot^2}
{\left[k_\bot^2 -M_{42}^2- i \varepsilon\right]
\left[k_\bot^2 -i \varepsilon\right]
}\,,
\nonumber\\
\textrm{~Res}_{k_4^-} f_2 &=& \frac{1}{x+\xi}\cdot\frac{y-a_4}{2(a_4-a_2)(a_4-a_3)} \cdot
\frac{\alpha_4 + \beta_4 k_\bot^2}
{\left[k_\bot^2 -M_{42}^2- i \varepsilon\right]
\left[k_\bot^2 - M_{43}^2-i \varepsilon\right]
}\,,
\label{eq:Residuab}
\end{eqnarray}
where $\alpha_i, \beta_i$  and $M_{ij}$ are now defined by:
\begin{eqnarray}
\alpha_{1,2} +\beta_{1,2} k_\bot^2  
&\equiv &
\mathcal{A}~x+\mathcal{B}\bigg|_{k^-_{1,2}} \,,
%\nonumber \\
%\alpha_2 +\beta_2 k_\bot^2  
%&\equiv &
%\mathcal{A}~x+\mathcal{B}\bigg|_{k^-_2} 
\nonumber \\
\alpha_{3,4} +\beta_{3,4} k_\bot^2  
&\equiv &
\mathcal{A}~\xi-\mathcal{B}\bigg|_{k^-_{2,4}} \,,
\nonumber \\
%\alpha_4 +\beta_4 k_\bot^2  
%&\equiv &
%\mathcal{A}~\xi-\mathcal{B}\bigg|_{k^-_4} 
%\nonumber \\
M_{ij}&\equiv & Q^2 \frac{(y-a_i)(a_j-y)}{x_B(a_i-a_j)}\,. 
\label{eq:alphabeta_defb}
\end{eqnarray}
Making use of Eq. (\ref{eq:ktinegrb}) we arrive to:
\begin{eqnarray}
I_{(5)} &=& -\frac{i}{(4\pi)^2} \left(\frac{1}{4\pi\mu^2}\right)^{\frac{\epsilon}{2}}\Gamma\left(-\frac{\epsilon}{2}\right)\frac{1}{x+\xi}\bigg\{
\nonumber\\
& &
\int_{a_1}^{a_3} dy \frac{y-a_1}{(a_1-a_2)(a_1-a_3)} 
\left[
\frac{\alpha_1 + \beta_1 M_{12}^2}{M_{12}^2-M_{13}^2} 
\left(-M_{12}^2-i\varepsilon \right)^{\frac{\epsilon}{2}}
-
\frac{\alpha_1 + \beta_1 M_{13}^2}{M_{12}^2-M_{13}^2} 
\left(-M_{13}^2-i\varepsilon \right)^{\frac{\epsilon}{2}}
\right]
\nonumber \\
%%%%%%%%%%%%%%%%%%%%%%%%%%%%%%%%%%%%%%%%%%%%%%%%%%%%%%%%%%%%%%%%%%%%%%%%%%%%%
&+&
\int_{a_2}^{a_3} dy \frac{y-a_2}{(a_2-a_1)(a_2-a_3)} 
\left[
\frac{\alpha_2 + \beta_2 M_{12}^2}{M_{12}^2} 
\left(-M_{12}^2-i\varepsilon \right)^{\frac{\epsilon}{2}}
\right]
\nonumber \\
%%%%%%%%%%%%%%%%%%%%%%%%%%%%%%%%%%%%%%%%%%%%%%%%%%%%%%%%%%%%%%%%%%%%%%%%%%%%%
%%%%%%%%%%%%%%%%%%%%%%%%%%%%%%%%%%%%%%%%%%%%%%%%%%%%%%%%%%%%%%%%%%%%%%%%%%%%%
%%%%%%%%%%%%%%%%%%%%%%%%%%%%%%%%%%%%%%%%%%%%%%%%%%%%%%%%%%%%%%%%%%%%%%%%%%%%%
&+&
\int_{a_4}^{a_3} dy \frac{y-a_4}{(a_4-a_2)(a_4-a_3)} 
\left[
\frac{\alpha_4 + \beta_4 M_{42}^2}{M_{42}^2-M_{43}^2} 
\left(-M_{42}^2-i\varepsilon \right)^{\frac{\epsilon}{2}}
-
\frac{\alpha_4 + \beta_4 M_{43}^2}{M_{42}^2-M_{43}^2} 
\left(-M_{43}^2-i\varepsilon \right)^{\frac{\epsilon}{2}}
\right]
\nonumber \\
%%%%%%%%%%%%%%%%%%%%%%%%%%%%%%%%%%%%%%%%%%%%%%%%%%%%%%%%%%%%%%%%%%%%%%%%%%%%%
&+&
\int_{a_2}^{a_3} dy \frac{y-a_2}{(a_2-a_4)(a_2-a_3)} 
\left[
\frac{\alpha_3 + \beta_3 M_{42}^2}{M_{42}^2} 
\left(-M_{42}^2-i\varepsilon \right)^{\frac{\epsilon}{2}}
\right] \bigg\} \,.
\label{eq:lastintegrationb}
\end{eqnarray}
One can check that the relations:
\begin{eqnarray}
\frac{y-a_1}{(a_1-a_2)(a_1-a_3)} 
\frac{1}{M_{12}^2-M_{13}^2}
&=&
\frac{x_B}{Q^2}\frac{1}{(a_3-a_2)(y-a_1)} \,,
\nonumber \\
\frac{y-a_2}{(a_2-a_1)(a_2-a_3)} 
\frac{1}{M_{12}^2}
&=&
\frac{x_B}{Q^2}\frac{1}{(a_2-a_3)(y-a_1)} \,,
\label{eq:relations}
\end{eqnarray}
%and $a_3-a_2 = x-\xi$, we get:
%\begin{eqnarray}
%I_{(5)} &=& -\frac{i}{(4\pi)^2} \left(\frac{1}{4\pi\mu^2}\right)^{\frac{\epsilon}{2}}\Gamma\left(-\frac{\epsilon}{2}\right)\frac{1}{x^2-\xi^2}\frac{x_B}{Q^2}\bigg\{
%\nonumber\\
%& &
%\int_{a_1}^{a_3} dy \frac{1}{y-a_1} 
%\left[
%(\alpha_1 + \beta_1 M_{12}^2) 
%\left(-M_{12}^2-i\varepsilon \right)^{\frac{\epsilon}{2}}
%-
%(\alpha_1 + \beta_1 M_{13}^2)
%\left(-M_{13}^2-i\varepsilon \right)^{\frac{\epsilon}{2}}
%\right]
%\nonumber \\
%%%%%%%%%%%%%%%%%%%%%%%%%%%%%%%%%%%%%%%%%%%%%%%%%%%%%%%%%%%%%%%%%%%%%%%%%%%%%%
%&-&
%\int_{a_2}^{a_3} dy \frac{1}{y-a_1} 
%\left[
%(\alpha_2 + \beta_2 M_{12}^2) 
%\left(-M_{12}^2-i\varepsilon \right)^{\frac{\epsilon}{2}}
%\right]
%\nonumber \\
%%%%%%%%%%%%%%%%%%%%%%%%%%%%%%%%%%%%%%%%%%%%%%%%%%%%%%%%%%%%%%%%%%%%%%%%%%%%%%
%%%%%%%%%%%%%%%%%%%%%%%%%%%%%%%%%%%%%%%%%%%%%%%%%%%%%%%%%%%%%%%%%%%%%%%%%%%%%%
%%%%%%%%%%%%%%%%%%%%%%%%%%%%%%%%%%%%%%%%%%%%%%%%%%%%%%%%%%%%%%%%%%%%%%%%%%%%%%
%&+&
%\int_{a_4}^{a_3} dy \frac{1}{y-a_4} 
%\left[
%(\alpha_4 + \beta_4 M_{42}^2) 
%\left(-M_{42}^2-i\varepsilon \right)^{\frac{\epsilon}{2}}
%-
%(\alpha_4 + \beta_4 M_{43}^2)
%\left(-M_{43}^2-i\varepsilon \right)^{\frac{\epsilon}{2}}
%\right]
%\nonumber \\
%%%%%%%%%%%%%%%%%%%%%%%%%%%%%%%%%%%%%%%%%%%%%%%%%%%%%%%%%%%%%%%%%%%%%%%%%%%%%%
%&-&
%\int_{a_2}^{a_3} dy \frac{1}{y-a_4} 
%\left[
%(\alpha_3 + \beta_3 M_{42}^2)
%\left(-M_{42}^2-i\varepsilon \right)^{\frac{\epsilon}{2}}
%\right] \bigg\}
%\label{eq:lastintegrationb2}
%\end{eqnarray}
%$\alpha_i + \beta_i M^2$ are awfull, but with Mathematica one can check that:
\begin{eqnarray}
\alpha_1 + \beta_1 M_{12}^2 &=& \alpha_2 + \beta_2 M_{12}^2 \,, \nonumber\\
\alpha_4 + \beta_4 M_{42}^2 &=& \alpha_3 + \beta_3 M_{42}^2\,,
\label{eq:fact}
\end{eqnarray}
allow to rearrange the integration limits, so we can express our integrals in the form allowing us to perform integration over $y$ again using the beta function definition:
\begin{eqnarray}
I_{(5)} &=& -\frac{i}{(4\pi)^2} \left(\frac{1}{4\pi\mu^2}\right)^{\frac{\epsilon}{2}}\Gamma\left(-\frac{\epsilon}{2}\right)\frac{1}{x^2-\xi^2}\frac{x_B}{Q^2}\bigg\{
\nonumber\\
& &
\int_{a_1}^{a_2} dy \frac{1}{y-a_1} 
\left[
(\alpha_1 + \beta_1 M_{12}^2) 
\left(-M_{12}^2-i\varepsilon \right)^{\frac{\epsilon}{2}}
\right]
\nonumber \\
%%%%%%%%%%%%%%%%%%%%%%%%%%%%%%%%%%%%%%%%%%%%%%%%%%%%%%%%%%%%%%%%%%%%%%%%%%%%%
&-&
\int_{a_1}^{a_3} dy \frac{1}{y-a_1} 
\left[
(\alpha_1 + \beta_1 M_{13}^2) 
\left(-M_{13}^2-i\varepsilon \right)^{\frac{\epsilon}{2}}
\right]
\nonumber \\
%%%%%%%%%%%%%%%%%%%%%%%%%%%%%%%%%%%%%%%%%%%%%%%%%%%%%%%%%%%%%%%%%%%%%%%%%%%%%
%%%%%%%%%%%%%%%%%%%%%%%%%%%%%%%%%%%%%%%%%%%%%%%%%%%%%%%%%%%%%%%%%%%%%%%%%%%%%
%%%%%%%%%%%%%%%%%%%%%%%%%%%%%%%%%%%%%%%%%%%%%%%%%%%%%%%%%%%%%%%%%%%%%%%%%%%%%
&+&
\int_{a_4}^{a_2} dy \frac{1}{y-a_4} 
\left[
(\alpha_4 + \beta_4 M_{42}^2) 
\left(-M_{42}^2-i\varepsilon \right)^{\frac{\epsilon}{2}}
\right]
\nonumber \\
%%%%%%%%%%%%%%%%%%%%%%%%%%%%%%%%%%%%%%%%%%%%%%%%%%%%%%%%%%%%%%%%%%%%%%%%%%%%%
&-&
\int_{a_4}^{a_3} dy \frac{1}{y-a_4} 
\left[
(\alpha_4 + \beta_4 M_{43}^2)
\left(-M_{43}^2-i\varepsilon \right)^{\frac{\epsilon}{2}}
\right] \bigg\} \,.
\label{eq:lastintegrationb3}
\end{eqnarray}
In the next section we will write explicitly the final results of all of the above calculations.
%----------------------------------------------------------------------
\section{Results}
We see that hard scattering amplitudes for general kinematics 
%(which are explicitely written in appendix \ref{app:HSA}) 
have the following structure:
\begin{eqnarray}
& & \frac{e^2\alpha_s C_F}{4\pi}\frac{1}{(4 \pi \mu^2)^{\frac{\epsilon}{2}}}
\bigg\{ \phantom{+}\left(Q^2\frac{x_B-x}{x_B}-i\varepsilon\right)^{\frac{\epsilon}{2}}\cdot f(x,\xi,x_B,\epsilon,\varepsilon)
\nonumber\\
& &\phantom{-i g_T^{\mu\nu}\frac{e^2\alpha_s C_F}{4\pi}\frac{1}{(4 \pi \mu^2)^{\frac{\epsilon}{2}}}\bigg\{}
+\left(Q^2\frac{x_B-\xi}{x_B}-i\varepsilon\right)^{\frac{\epsilon}{2}}\cdot g(x,\xi,x_B,\epsilon,\varepsilon)
\bigg\} + (x_B \longleftrightarrow -x_B) \,,
\label{eq:Gen_structure}
\end{eqnarray}
which in the $\epsilon \to 0$ limit for $q\gamma \to q\gamma$ amplitude gives:
\begin{eqnarray}
&&\tr\left[\mathcal{M}^{\mu\nu}\not\!{p}\right]
= g_T^{\mu\nu}\frac{e^2\alpha_s C_F}{4\pi}\left(\frac{|Q^2|}{4 \pi \mu^2}\right)^{\frac{\epsilon}{2}}
\bigg\{
\nonumber \\
&& 
\frac{1}{\epsilon}\bigg[\frac{12}{x-x_B+i\varepsilon\frac{x_B}{Q^2}}
\nonumber \\
&&
+\left(
\frac{16(xx_B -\xi^2)}{(x-x_B+i\varepsilon\frac{x_B}{Q^2})(x^2-\xi^2)}
+\frac{8(x-x_B)}{x^2-\xi^2}
\right)
\log \left(sgn(Q^2)\frac{x_B-x}{x_B}-i\varepsilon\right)
\nonumber \\
&&
+
\left(
\frac{8(\xi -x_B)}{(x-x_B+i\varepsilon\frac{x_B}{Q^2})(x-\xi)}
-
\frac{8(\xi -x_B)}{(x+x_B-i\epsilon\frac{x_B}{Q^2})(x+\xi)}
+\frac{8x(x_B-\xi)}{\xi(x^2-\xi^2)}
\right)
\log \left(sgn(Q^2)\frac{x_B-\xi}{x_B}-i\varepsilon\right) \bigg]
\nonumber \\
&& - \frac{18}{x-x_B+i\varepsilon\frac{x_B}{Q^2}} 
\nonumber\\
&&
+ 6\frac{x^2+\xi^2-2xx_B}{(x-x_B+i\varepsilon\frac{x_B}{Q^2})(x^2-\xi^2)}
\log \left(sgn(Q^2)\frac{x_B-x}{x_B}-i\varepsilon\right) 
\nonumber\\
&&
+\left(
\frac{4(xx_B -\xi^2)}{(x-x_B+i\varepsilon\frac{x_B}{Q^2})(x^2-\xi^2)}
+\frac{2(x-x_B)}{x^2-\xi^2}
\right)
\log ^2\left(sgn(Q^2)\frac{x_B-x}{x_B}-i\varepsilon\right)
\nonumber \\
&&
+
\left(
\frac{6(\xi -x_B)}{(x+x_B-i\epsilon\frac{x_B}{Q^2})(x+\xi)}
-
\frac{6(\xi -x_B)}{(x-x_B+i\varepsilon\frac{x_B}{Q^2})(x-\xi)}
\right)
\log \left(sgn(Q^2)\frac{x_B-\xi}{x_B}-i\varepsilon\right)
\nonumber \\
&&
+
\left(
\frac{2(\xi -x_B)}{(x-x_B+i\varepsilon\frac{x_B}{Q^2})(x-\xi)}
-
\frac{2(\xi -x_B)}{(x+x_B-i\epsilon\frac{x_B}{Q^2})(x+\xi)}
+\frac{2x(x_B-\xi)}{\xi(x^2-\xi^2)}
\right)
\log^2 \left(sgn(Q^2)\frac{x_B-\xi}{x_B}-i\varepsilon\right) \bigg\} 
\nonumber \\
&+& 
 (x_B \longleftrightarrow -x_B) \,,
\label{eq:qgenampl}
\end{eqnarray}
and for $g \gamma \to g \gamma$:
\begin{eqnarray}
g_T^{\mu\nu}g_T^{\alpha\beta}\mathcal{M}_{\mu\nu\alpha\beta}
&=& \frac{e^2\alpha_s T_F}{4\pi}\left(\frac{|Q^2|}{4 \pi \mu^2}\right)^{\frac{\epsilon}{2}}
\bigg\{
\nonumber \\
&& 
\frac{1}{\epsilon}\bigg[
-\frac{32(x^2-2x_Bx+2x_B^2 -\xi^2)}{x^2-\xi^2}
\log \left(sgn(Q^2)\frac{x_B-x}{x_B}-i\varepsilon\right)
\nonumber \\
&&
-\frac{32(x_B-\xi)(x^2-2x_B\xi -\xi^2)}{\xi(x^2-\xi^2)}
\log \left(sgn(Q^2)\frac{x_B-\xi}{x_B}-i\varepsilon\right) \bigg]
\nonumber \\
&&
- 16
\log \left(sgn(Q^2)\frac{x_B-x}{x_B}-i\varepsilon\right) 
\nonumber\\
&&
-\frac{8(x^2-2x_Bx+2x_B^2 -\xi^2)}{x^2-\xi^2}
\log ^2\left(sgn(Q^2)\frac{x_B-x}{x_B}-i\varepsilon\right)
\nonumber \\
&&
+
16\left(
1-\frac{x_B}{\xi}
\right)
\log \left(sgn(Q^2)\frac{x_B-\xi}{x_B}-i\varepsilon\right)
\nonumber \\
&&
-\frac{8(x_B-\xi)(x^2-2x_B\xi -\xi^2)}{\xi(x^2-\xi^2)}\log^2 \left(sgn(Q^2)\frac{x_B-\xi}{x_B}-i\varepsilon\right) 
\bigg\} 
\nonumber \\
&+&  (x_B \longleftrightarrow -x_B) \,.
\label{eq:ggenampl}
\end{eqnarray}
From the above result one can easily read the coefficient functions defined by Eq. (\ref{eq:coeff}), necessary for a calculation of the DDVCS amplitude, by means of Eq. (\ref{eq:ceofficients}). However, most experimental data is, or will be available for either DVCS or TCS, so we will elaborate more on those cases.  

Below we present the resulting coefficient functions for the limiting cases of DVCS ($Q^2 >0$ and $x_B = \xi$) and TCS ($-Q^2 =Q'^2 >0$ and $x_B = -\xi$).
\subsection{DVCS limit}
%%%%%%%%%%%%%%%%%%%%%%%%%%%%%%%%%%%%%%%%%%%%%%%%%%%%%%%%%%%%%%%%%%%%%%%%%%%%%%%%
We present the results explicitly showing $i\varepsilon$ terms that uniquely determine all imaginary parts. Quark coefficient functions read:
\begin{eqnarray}
C_0^q &=& e_q^2\left(\frac{1}{x-\xi+i\varepsilon}+\frac{1}{x+\xi-i\varepsilon}\right) \,, \nonumber \\
C_1^q &=& \frac{e_q^2\alpha_SC_F}{4\pi}
\bigg\{
\frac{1}{x-\xi+i\varepsilon}
\bigg[
-9+3\log(1-\frac{x}{\xi}-i\varepsilon)
-6\frac{\xi}{x+\xi}\log(1-\frac{x}{\xi}-i\varepsilon)
+6\frac{\xi}{x+\xi}\log 2
\nonumber \\ && \phantom{AAAAAAAAAAAAAAA}
+\log^2(1-\frac{x}{\xi}-i\varepsilon )-\log^22
\bigg]\nonumber\\
&& 
\phantom{\frac{e_q^2\alpha_SC_F}{4\pi}}+\frac{1}{x+\xi-i\varepsilon}
\bigg[
-9+3\log(1+\frac{x}{\xi}-i\varepsilon)
+6\frac{\xi}{x-\xi}\log(1+\frac{x}{\xi}-i\varepsilon)
-6\frac{\xi}{x-\xi}\log 2
\nonumber \\ && \phantom{AAAAAAAAAAAAAAA}
+\log^2(1+\frac{x}{\xi}-i\varepsilon )-\log^22
\bigg]
\bigg\} \,, \nonumber\\
C_{coll}^q &=&\frac{e_q^2\alpha_SC_F}{4\pi}
\bigg\{
\frac{1}{x-\xi+i\varepsilon}
\bigg[
6+4\log(1-\frac{x}{\xi}-i\varepsilon) - 4 \log2
\bigg]\nonumber\\
&& \phantom{\frac{e_q^2\alpha_SC_F}{4\pi}}
+
\frac{1}{x+\xi-i\varepsilon}
\bigg[
6+4\log(1+\frac{x}{\xi}-i\varepsilon) - 4 \log2
\bigg]\bigg\} \,.
\label{eq:QDVCS}
\end{eqnarray}
%%%%%%%%%%%%%%%%%%%%%%%%%%%%%%%%%%%%%%%%%%%%%%%%%%%%%%%%%%%%%%%%%%%%%%%%%%%%%%%%
Gluon coefficient functions read:
\begin{eqnarray}
C_{coll}^g &=& \frac{\left(\sum_qe_q^2\right)\alpha_ST_F}{4\pi}\frac{8x}{(x+\xi-i\varepsilon)(x-\xi+i\varepsilon)}\cdot \nonumber \\
&&\left[
\frac{x-\xi}{x+\xi}\log\left(1-\frac{x}{\xi} - i \varepsilon\right)
+\frac{x+\xi}{x-\xi}\log\left(1+\frac{x}{\xi} - i \varepsilon\right)
-2\frac{x^2+\xi^2}{x^2-\xi^2}\log 2 \right]\,,\nonumber\\
C_1^g &=& \frac{\left(\sum_qe_q^2\right)\alpha_ST_F}{4\pi}\frac{2x}{(x+\xi-i\varepsilon)(x-\xi+i\varepsilon)}\cdot \nonumber \\
&&\bigg[
-2\frac{x-3\xi}{x+\xi}\log\left(1-\frac{x}{\xi} - i \varepsilon\right)
+\frac{x-\xi}{x+\xi}\log^2\left(1-\frac{x}{\xi} - i \varepsilon\right)
\nonumber\\&&\phantom{\bigg[}
-2\frac{x+3\xi}{x-\xi}\log\left(1+\frac{x}{\xi} - i \varepsilon\right)
+\frac{x+\xi}{x-\xi}\log^2\left(1+\frac{x}{\xi} - i \varepsilon\right)
\nonumber\\&&\phantom{\bigg[}
+4\frac{x^2+3\xi^2}{x^2-\xi^2}\log 2 
-2\frac{x^2+\xi^2}{x^2-\xi^2}\log^2 2 
\bigg] \,.
\label{eq:GDVCS}
\end{eqnarray}
Although the result for $C_1^g$ contains dangerously looking denominators inside the square parenthesis, it is easy to check that the expression inside those parenthesis is regular in the limits $x \to \pm\xi$.

Above results are in agreement with earlier results  \cite{JiO,Mankiewicz} which were obtained in an unphysical region of parameter space, and then analytically continued to obtain DVCS case. We see that the simple prescripton that all imaginary parts can be obtained by substracting a small imaginary part from $\xi$, i.e. $\xi \to \xi-i\varepsilon$, is confirmed by our calculations. 
\subsection{TCS limit}
Quark coefficient functions read: 
%%%%%%%%%%%%%%%%%%%%%%%%%%%%%%%%%%%%%%%%%%%%%%%%%%%%%%%%%%%%%%%%%%%%%%%%%%%%%%%%
\begin{eqnarray}
C_0^q &=& e_q^2\left(\frac{1}{x-\xi-i\varepsilon}+\frac{1}{x+\xi+i\varepsilon}\right) \,,\nonumber \\
C_1^q &=& \frac{e_q^2\alpha_SC_F}{4\pi}
\bigg\{
\frac{1}{x-\xi-i\varepsilon}
\bigg[
-9+3\log(-1+\frac{x}{\xi}-i\varepsilon)
-6\frac{\xi}{x+\xi}\log(-1+\frac{x}{\xi}-i\varepsilon)
+6\frac{\xi}{x+\xi}\log (-2-i\varepsilon)
\nonumber \\ && \phantom{AAAAAAAAAAAAAAA}
+\log^2(-1+\frac{x}{\xi}-i\varepsilon )-\log^2(-2-i\varepsilon)
\bigg]\nonumber\\
&& 
\phantom{\frac{e_q^2\alpha_SC_F}{4\pi}}+\frac{1}{x+\xi+i\varepsilon}
\bigg[
-9+3\log(-1-\frac{x}{\xi}-i\varepsilon)
+6\frac{\xi}{x-\xi}\log(-1-\frac{x}{\xi}-i\varepsilon)
-6\frac{\xi}{x-\xi}\log (-2-i\varepsilon)
\nonumber \\ && \phantom{AAAAAAAAAAAAAAA}
+\log^2(-1-\frac{x}{\xi}-i\varepsilon )-\log^2(-2-i\varepsilon)
\bigg]
\bigg\} \,,\nonumber\\
C_{coll}^q &=&\frac{e_q^2\alpha_SC_F}{4\pi}
\bigg\{
\frac{1}{x-\xi-i\varepsilon}
\bigg[
6+4\log(-1+\frac{x}{\xi}-i\varepsilon) - 4 \log(-2-i\varepsilon)
\bigg]\nonumber\\
&& \phantom{\frac{e_q^2\alpha_SC_F}{4\pi}}
+
\frac{1}{x+\xi+i\varepsilon}
\bigg[
6+4\log(-1-\frac{x}{\xi}-i\varepsilon) - 4\log(-2-i\varepsilon)
\bigg]\bigg\}\,.
\label{eq:QTCS}
\end{eqnarray}
%%%%%%%%%%%%%%%%%%%%%%%%%%%%%%%%%%%%%%%%%%%%%%%%%%%%%%%%%%%%%%%%%%%%%%%%%%%%%%%%
Gluon coefficient functions read:
\begin{eqnarray}
C_{coll}^g &=& \frac{\left(\sum_qe_q^2\right)\alpha_ST_F}{4\pi}\frac{8x}{(x+\xi+i\varepsilon)(x-\xi-i\varepsilon)}\cdot \nonumber \\
&&\left[
\frac{x-\xi}{x+\xi}\log\left(-1+\frac{x}{\xi} - i \varepsilon\right)
+\frac{x+\xi}{x-\xi}\log\left(-1-\frac{x}{\xi} - i \varepsilon\right)
-2\frac{x^2+\xi^2}{x^2-\xi^2}\log(-2-i\varepsilon) \right] \,,\nonumber\\
C_1^g &=& \frac{\left(\sum_qe_q^2\right)\alpha_ST_F}{4\pi}\frac{2x}{(x+\xi+i\varepsilon)(x-\xi-i\varepsilon)}\cdot \nonumber \\
&&\bigg[
-2\frac{x-3\xi}{x+\xi}\log\left(-1+\frac{x}{\xi} - i \varepsilon\right)
+\frac{x-\xi}{x+\xi}\log^2\left(-1+\frac{x}{\xi} - i \varepsilon\right)
\nonumber\\&&\phantom{\bigg[}
-2\frac{x+3\xi}{x-\xi}\log\left(-1-\frac{x}{\xi} - i \varepsilon\right)
+\frac{x+\xi}{x-\xi}\log^2\left(-1-\frac{x}{\xi} - i \varepsilon\right)
\nonumber\\&&\phantom{\bigg[}
+4\frac{x^2+3\xi^2}{x^2-\xi^2}\log (-2-i\varepsilon) 
-2\frac{x^2+\xi^2}{x^2-\xi^2}\log^2 (-2-i\varepsilon) 
\bigg]\,.
\label{eq:GTCS}
\end{eqnarray}
As in the DVCS case terms inside the square parenthesis of $C_1^g$ are regular in the limits $x \to \pm\xi$.

There are some important differences between the Eqs. (\ref{eq:QTCS}, \ref{eq:GTCS}) describing the TCS case and  Eqs. (\ref{eq:QDVCS}, \ref{eq:GDVCS}) describing DVCS. First we notice that we have to add small imaginary part to $\xi$, i.e. $\xi \to \xi+i\varepsilon$, rather then substract as in the DVCS case. The second difference is the minus sign under the logarithms, which produces additional terms. Particularly $\log^2(-2-i\varepsilon)$ present in the TCS result may produce correction much bigger then the $\log^2(2)$ in the DVCS case. Another important difference between the DVCS and TCS amplitudes concerns their imaginary parts, which in the DVCS case is present only in the DGLAP region, while in the TCS case, it is present in both DGLAP and ERBL regions.

\section{Conclusions} 
It is well known that at the Born level TCS and DVCS hard scattering amplitudes are related: 
\begin{eqnarray}
C_{0(DVCS)}^q = {C_{0(TCS)}^q}^*\,.
\label{eq:BornTCSDVCS}
\end{eqnarray}
The same relation holds for the collinear terms:
\begin{eqnarray}
C_{coll(DVCS)}^q = {C_{coll(TCS)}^q}^*\,,
\label{eq:collTCSDVCS}
\end{eqnarray}
as they are equal to the convolution of the same evolution kernel with Born level amplitudes. Indeed this equality is crucial for factorization to hold. But in the NLO this relation no longer holds. For the quark part, we have :
\begin{eqnarray}
\frac{{C_{1(TCS)}^q}^* - C_{1(DVCS)}^q}{\frac{e^2\alpha_SC_F}{4\pi}}  & = &
\frac{1}{x-\xi+i\varepsilon}\left[\left(3-2\log 2 +2\log|1-\frac{x}{\xi}|\right)(i\pi)
+\pi^2\left(1+\theta(x-\xi)-\theta(-x+\xi\right))\right]\nonumber\\
&+&\frac{1}{x+\xi-i\varepsilon}\left[\left(3-2\log 2 +2\log|1+\frac{x}{\xi}|\right)(i\pi)
+\pi^2\left(1+\theta(-x-\xi)-\theta(x+\xi\right))\right]
\label{eq:NLOTCSDVCS}
\end{eqnarray}
%%%%%%%%%%%%%%%%%%%%%%%%%%%%%%%%% FIGURE
\begin{figure}[t]
\begin{center}
\rput(3.6,4.4){TCS, ERBL}
\rput(12,4.4){TCS, DGLAP}
\rput(3.6,-0.5){DVCS, ERBL}
\rput(12,-0.5){DVCS, DGLAP}
\epsfxsize=0.39\textwidth
\epsffile{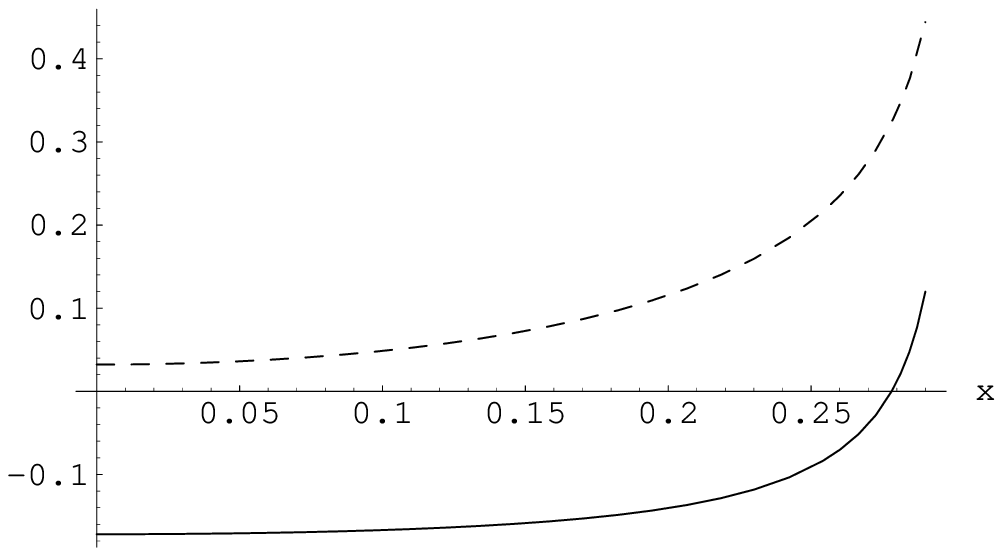}
\hspace{0.05\textwidth}
\epsfxsize=0.39\textwidth
\epsffile{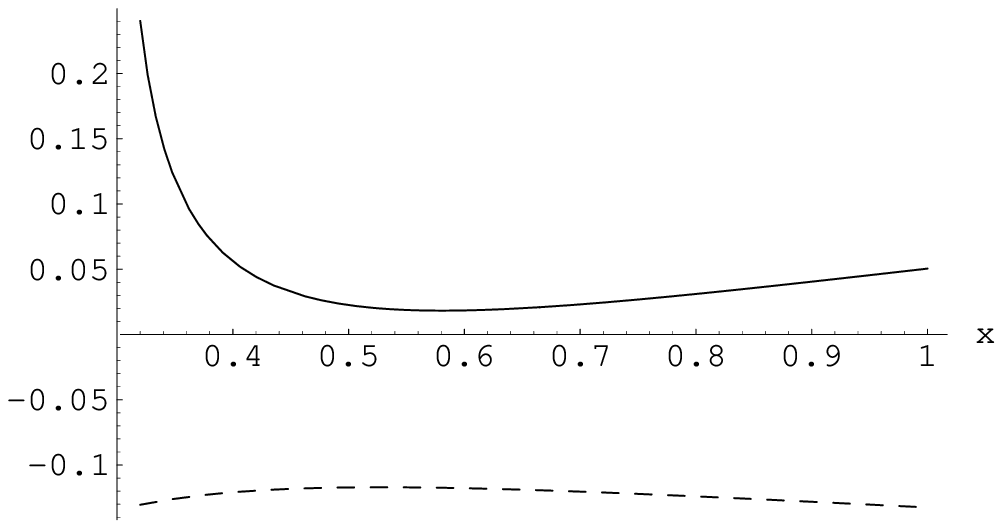}
\vspace{0.04\textwidth}

\epsfxsize=0.39\textwidth
\epsffile{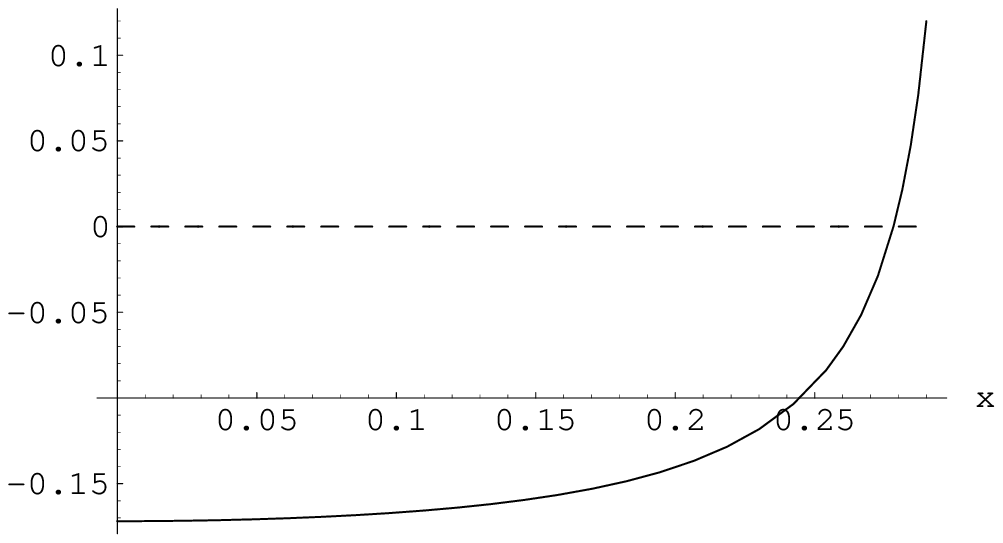}
\hspace{0.05\textwidth}
\epsfxsize=0.39\textwidth
\epsffile{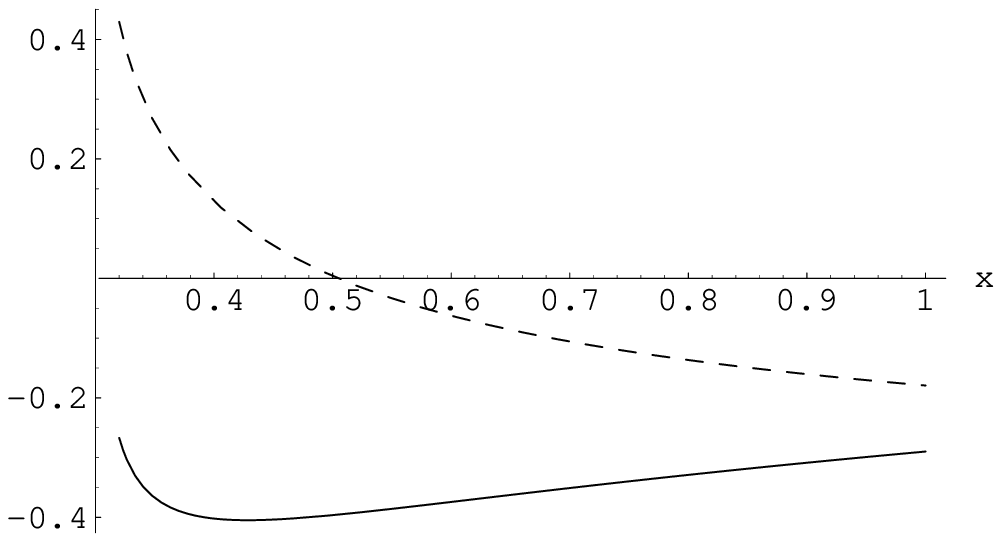}
\caption{\label{Fig:ratio} Real (solid line) and imaginary (dashed line) part of the ratio $R^q$ of the NLO quark coefficient function to the Born term in Timelike Compton Scattering (up) and Deeply Virtual Compton Scattering (down) as a function of $x$ in the ERBL (left) and DGLAP (right) region for $\xi = 0.3$, for $\mu_F^2 = |Q^2|$. }
\end{center}
\end{figure}
%%%%%%%%%%%%%%%%%%%%%%%%%%%%%%%%%%%%%%%%%%%%
To discuss this difference and present the magnitude of corrections we define the following ratio:
\begin{eqnarray}
R^q = \frac{C_{1}^q+\frac{1}{2}\log \left(\frac{|Q^2|}{\mu_F^2}\right) \cdot C^q_{coll}}{C^q_{0}}
\label{eq:r}
\end{eqnarray}
of the NLO quark correction to the coefficient function, to the Born level.
On Fig. \ref{Fig:ratio} we  show for $\mu_F^2 = |Q^2| $ the real and imaginary parts of the ratio $R^q$ in timelike and spacelike Compton Scattering as a function of $x$ in the ERBL (left) and DGLAP (right) region for $\xi = 0.3$. We  fix $\alpha_s = 0.25$ and restrict the plots to the positive $x$ region, as the coefficient functions are antisymmetric in that variable. We see that in the TCS case, the imaginary part of the amplitude is  present in both the ERBL and DGLAP regions, contrarily to the DVCS case, where it exists  only in the DGLAP region. The magnitude of these NLO coefficient functions is not negligible. We see that the importance of these NLO coefficient functions is magnified when we consider the  difference of the coefficient functions  ${C_{1(TCS)}^q}^* - C_{1(DVCS)}^q  $. The conclusion is that extracting the universal GPDs from both TCS and DVCS reactions requires much care. 

%%%%%%%%%%%%%%%%%%%%%%%%%%%%%%%%% FIGURE
\begin{figure}[t]
\begin{center}
\epsfxsize=0.5\textwidth
\epsffile{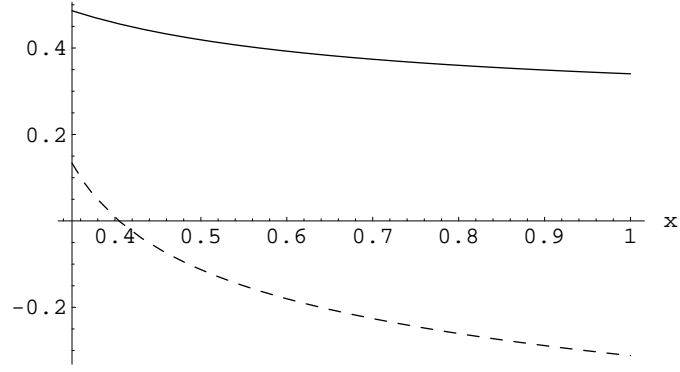}
\caption{\label{Fig:diff} 
Real (solid line) and imaginary (dashed line) part of the ratio $R^q_{T-S}$ of difference of NLO quark coefficient functions to the LO coefficient functions in the TCS and DVCS as a function of $x$ in the DGLAP  region for $\xi = 0.3$.}
\end{center}
\end{figure}
%%%%%%%%%%%%%%%%%%%%%%%%%%%%%%%%%%%%%%%%%%%%
As is well known in inclusive reactions, one may choose a renormalization scheme (named DIS scheme \cite{DIS}) defined by the fact that NLO corrections to some observables vanish. This of course does not preclude the importance of NNLO corrections. In the exclusive case, we thus may propose that NLO corrections vanish in the DVCS amplitude. This {\em DVCS factorization scheme} then transfers all NLO corrections calculated here to the TCS coefficient functions, which become very sizeable. We illustrate this fact by showing on Fig. \ref{Fig:diff} the ratio $R^q_{T-S}$ of the difference of NLO quark coefficient functions to the LO coefficient function 
\begin{eqnarray}
R_{T-S}^q = \frac{C_{1(TCS)}^q- C_{1(DVCS)}^{q\,*}}{C^q_{0}}\,.
\label{eq:diff}
\end{eqnarray}

A final word is needed with respect to the presence of the $\pi^2 $ terms in the  difference of the NLO coefficient functions. Quite a rich literature \cite{DIS,Kfactor} exists on the importance of such factors in inclusive coefficient functions and their relation to soft gluon exchange. One may verify that in the exclusive case that we study here, a soft gluon approximation gives some of the $\pi^2$ terms that one may read from Eq. (\ref{eq:NLOTCSDVCS}). One can suppose that these corrections exponentiate when all order corrections are summed up. A  particular feature is worth to be pointed out : these  $\pi^2$ terms only exist in the DGLAP regions. We confess that we do not understand why this is the case. 

%%%%%%%%%%%%%%%%%%%%%%%%%%%%%%%%% FIGURE
\begin{figure}[h!]
\begin{center}
\epsfxsize=0.5\textwidth
\epsffile{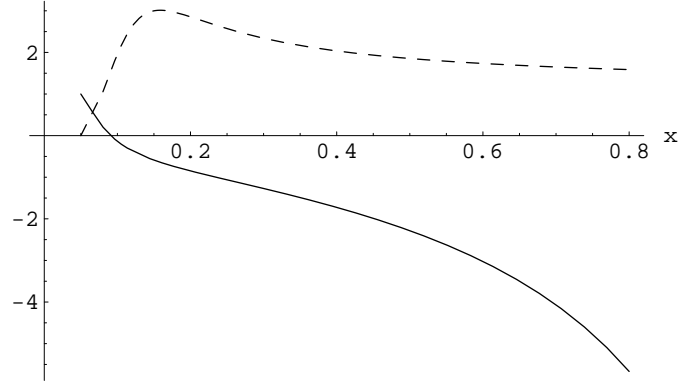}
\caption{\label{Fig:ratiog} 
Ratio of the real (solid line) and imaginary (dashed line) part of the NLO gluon coefficient function in TCS to the same quantity in DVCS as a function of $x$ in the DGLAP  region for $\xi = 0.05$ for $\mu_F^2 = |Q^2|$.}
\end{center}
\end{figure}
%%%%%%%%%%%%%%%%%%%%%%%%%%%%%%%%%%%%%%%%%%%%%%%

Let us now briefly comment on  the gluon coefficient functions. 
As in the case of quark corrections, the collinear parts are complex conjugated to each other:
\begin{eqnarray}
C_{coll(DVCS)}^g = {C_{coll(TCS)}^g}^*\,.
\label{eq:collgTCSDVCS}
\end{eqnarray}
Moreover,  the  real parts of the gluon contribution are equal for DVCS and TCS in the ERBL region. 
The differences between TCS and DVCS emerges in the ERBL region through the imaginary part of the coefficient function which is non zero only for the TCS case and is of the order of the real part. 
In the DGLAP region, the difference reads :
\begin{eqnarray}
\frac{C^g_{1(TCS)} - C^g_{1(DVCS)}}{\frac{\left(\sum_qe_q^2\right)\alpha_ST_F}{4\pi}} &\stackrel{x>\xi}{=}& \frac{2x}{x^2-\xi^2} 
\bigg[2 \frac{x-\xi}{x+\xi} \pi^2 \nonumber\\
&& \hspace*{-1cm}
+\left(-4\frac{x-3\xi}{x+\xi} +2\frac{x-\xi}{x+\xi}\log|1-\frac{x}{\xi}|-2\frac{x+\xi}{x-\xi}\log|1+\frac{x}{\xi}|
+4\frac{x^2+\xi^2}{x^2-\xi^2}\log 2 \right)(-i\pi) 
\bigg]\,,
\label{eq:gdiff}
\end{eqnarray} 
showing a sizeable difference  of the contributions to both the real and  imaginary parts of the amplitude.
In Fig. \ref{Fig:ratiog} we illustrate the ratio of the NLO gluon correction to the hard scattering amplitude in TCS to the same quantity in the DVCS in the DGLAP  region for $\xi = 0.05$ for $\mu_F^2 = |Q^2|$. 

\begin{figure}[h]
\begin{center}
\epsfxsize=0.39\textwidth
\epsffile{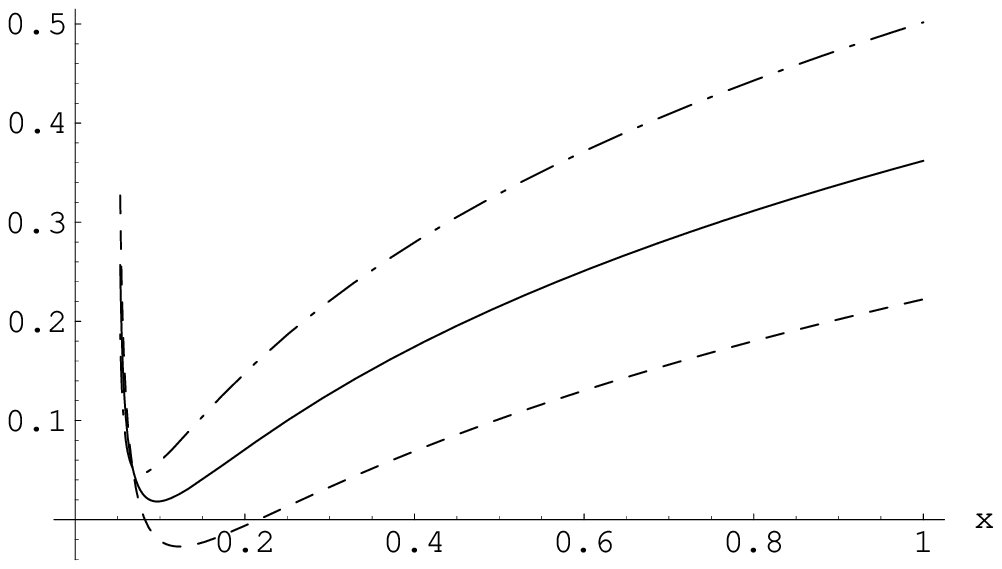}
\hspace{0.05\textwidth}
\epsfxsize=0.39\textwidth
\epsffile{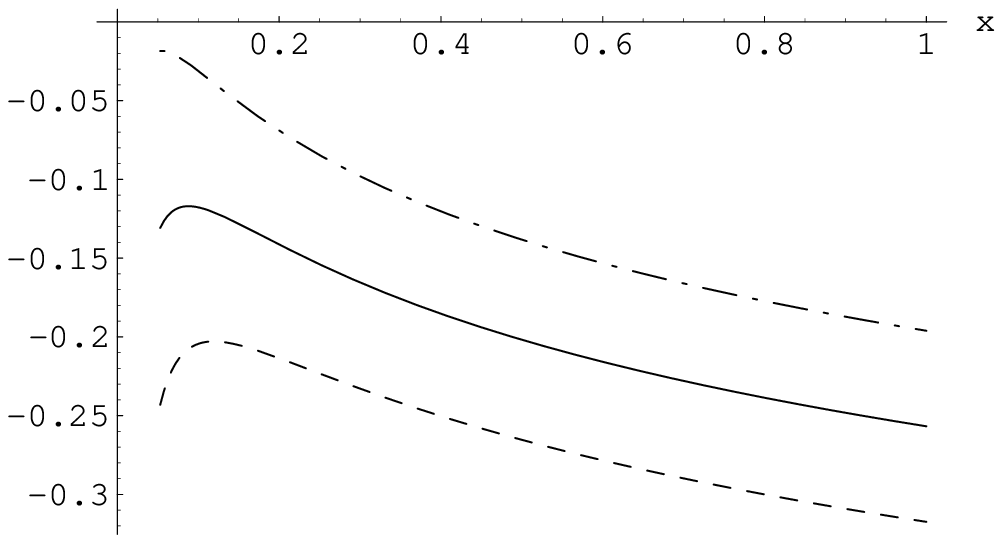}
\caption{Factorization scale dependence of the real (left) and imaginary (right) parts of ratio $R^q$ of NLO quark correction to hard scattering amplitudes to Born level coefficient function of the Timelike Compton Scattering as a function of $x$ in the DGLAP region for $\xi = 0.05$. The ratios are plotted for the values of $\frac{|Q^2|}{\mu_F^2}$ equal $0.5$ (dashed), $1$ (solid) and $2$ (dash-dotted line).}
\label{Fig:factscaledep}
\end{center}
\end{figure}

The discussion of NLO corrections to a hard scattering amplitude necessarily brings the question of the factorization scale dependence. On the Fig. \ref{Fig:factscaledep} we show the real and imaginary parts of the ratio $R^q$ of  NLO quark correction to hard scattering amplitudes to Born level coefficient function of the timelike Compton scattering as a function of $x$ in the DGLAP region for $\xi = 0.05$. The figures are plotted for various values of $\frac{|Q^2|}{\mu_F^2}$, and present a strong factorization scale dependence.

\begin{figure}[h]
\begin{center}
\epsfxsize=0.39\textwidth
\epsffile{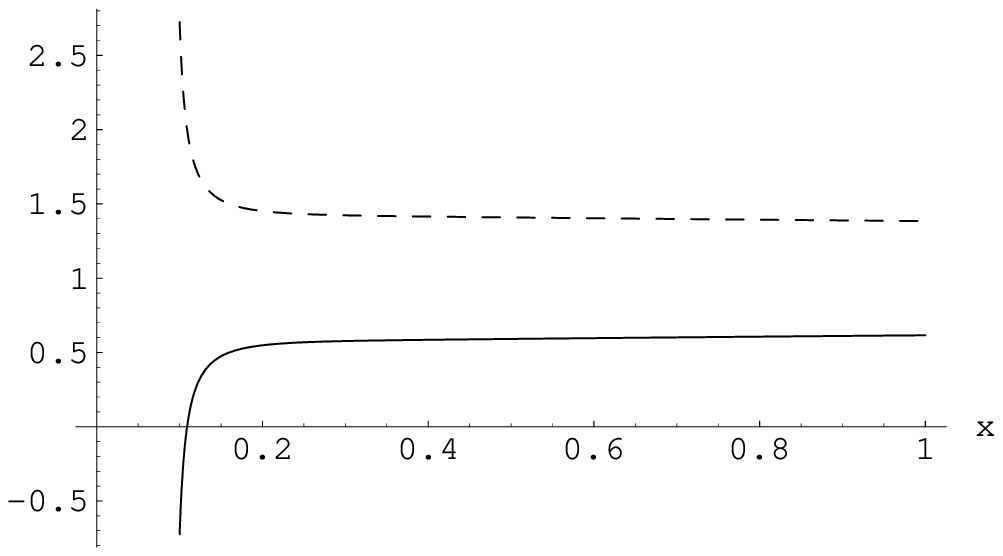}
\hspace{0.05\textwidth}
\epsfxsize=0.39\textwidth
\epsffile{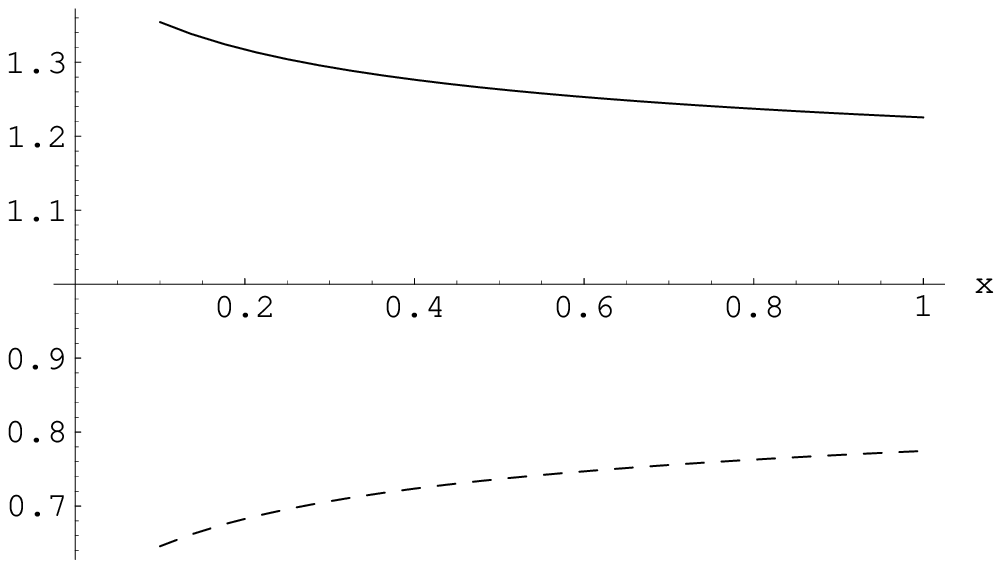}
\caption{Ratios of the real (left) and imaginary (right) parts of NLO gluon coefficient function for $|Q^2|=1/2 \mu_F^2 $ (solid line) and $|Q^2|= 2 \mu_F^2 $ (dashed line) to the same quantities with $|Q^2|= \mu_F^2$. Those quantities are calculated  for the timelike Compton scattering and plotted as a function of $x$ in the DGLAP region for $\xi = 0.05$.}
\label{Fig:factscaledepg}
\end{center}
\end{figure}

On the Fig. \ref{Fig:factscaledepg} we show the ratios of the real (left) and imaginary (right) parts of NLO gluon coefficient function for $|Q^2|=\frac{\mu_F^2}{2} $ (solid line) and $|Q^2|= 2 \mu_F^2 $ (dashed line) to the same quantities with $|Q^2|= \mu_F^2$. Those quantities are calculated  for the timelike Compton scattering and plotted as a function of $x$ in the DGLAP region for $\xi = 0.05$. Also in this case we notice a strong factorization scale dependence.
%%%%%%%%%%%%%%%%%%%%%%%%%%%%%%%%%%%%%%%%%%%%

Much phenomenological studies need now to be performed, by convoluting the coefficient functions to realistic GPDs and calculating the relevant observables. We plan to progress on these points in the near future.

%%%%%%%%%%%%%%%%%%%%%%%%%%%%%%%%% FIGURE
%%%%%%%%%%%%%%%%%%%%%%%%%%%%%%%%%%%%
%%%%%%%%%%%%%%%%%%%%%%%%%%%%%%%%% FIGURE
%%%%%%%%%%%%%%%%%%%%%%%%%%%%%%%%%%%%
%%%%%%%%%%%%%%%%%%%%%%%%%%%%%%%%%%%%
\section*{Acknowledgements} 
We are grateful to Igor Anikin, Markus Diehl, Michel Guidal, Dieter M\"{u}ller, Franck Sabati\'e, Jean Philippe Lansberg, Stepan Stepanyan, Pawel Nadel-Turonski and Samuel Wallon for useful discussions and correspondance. 
	
This work is partly supported by the Polish Grant N202 249235 
%%%%%%%%%%%%%%%%%%%%%%%%%%%%%%%%%%%%%%%%%%%%%%%%%%%%%%%%%%%%%%%%%%%%%%


\begin{thebibliography}{99}
%%%%%%%%%%%%%%%%%%%%%%%%%%%%%%%%%%%%%%%%%%%%%%%%%%%%%%%%%%%%%%%%%%%%%%
\bibitem{DVCSexp}
%\cite{Airapetian:2001yk}
%\bibitem{Airapetian:2001yk}
  A.~Airapetian {\it et al.}  [HERMES Collaboration],
  %``Measurement of the beam spin azimuthal asymmetry associated with
  %deeply-virtual Compton scattering,''
  Phys.\ Rev.\ Lett.\  {\bf 87}, 182001 (2001)
  [arXiv:hep-ex/0106068].
  %%CITATION = PRLTA,87,182001;%%

%\cite{Munoz Camacho:2006hx}
%\bibitem{Munoz Camacho:2006hx}
  C.~Munoz Camacho {\it et al.}  [Jefferson Lab Hall A Collaboration and Hall
                  A DVCS Collaboration],
  %``Scaling tests of the cross section for deeply virtual Compton
  %scattering,''
  Phys.\ Rev.\ Lett.\  {\bf 97} (2006) 262002
  [arXiv:nucl-ex/0607029].
  %%CITATION = PRLTA,97,262002;%%

%\cite{Chekanov:2003ya}
%\bibitem{Chekanov:2003ya}
  S.~Chekanov {\it et al.}  [ZEUS Collaboration],
  %``Measurement of deeply virtual Compton scattering at HERA,''
  Phys.\ Lett.\  B {\bf 573}, 46 (2003)
  [arXiv:hep-ex/0305028].
  %%CITATION = PHLTA,B573,46;%%

%\cite{Aktas:2005ty}
%\bibitem{Aktas:2005ty}
  A.~Aktas {\it et al.}  [H1 Collaboration],
  %``Measurement of deeply virtual Compton scattering at HERA,''
  Eur.\ Phys.\ J.\  C {\bf 44}, 1 (2005)
  [arXiv:hep-ex/0505061].
  %%CITATION = EPHJA,C44,1;%%

%\cite{Stepanyan:2001sm}
%\bibitem{Stepanyan:2001sm}
  S.~Stepanyan {\it et al.}  [CLAS Collaboration],
  %``First observation of exclusive deeply virtual Compton scattering in
  %polarized electron beam asymmetry measurements,''
  Phys.\ Rev.\ Lett.\  {\bf 87}, 182002 (2001)
  [arXiv:hep-ex/0107043].
  %%CITATION = PRLTA,87,182002;%%

%\cite{Kumericki:2007sa}
\bibitem{fitting}
%\bibitem{Kumericki:2007sa}
  K.~Kumericki, D.~Mueller and K.~Passek-Kumericki,
  %``Towards a fitting procedure for deeply virtual Compton scattering at
  %next-to-leading order and beyond,''
  Nucl.\ Phys.\  B {\bf 794}, 244 (2008)
  [arXiv:hep-ph/0703179].
  %%CITATION = NUPHA,B794,244;%%  

%\cite{Guidal:2009aa}
%\bibitem{Guidal:2009aa}
  M.~Guidal and H.~Moutarde,
  %``Generalized Parton Distributions from Deeply Virtual Compton Scattering at
  %HERMES,''
  Eur.\ Phys.\ J.\  A {\bf 42}, 71 (2009)
  [arXiv:0905.1220 [hep-ph]].
  %%CITATION = EPHJA,A42,71;%%

%\cite{Moutarde:2010uk}
%\bibitem{Moutarde:2010uk}
  H.~Moutarde,
  %``Extraction of the Compton Form Factor H from DVCS Measurements in the Quark
  %Sector,''
  arXiv:1010.4521 [hep-ph].
  %%CITATION = ARXIV:1010.4521;%%

%\cite{Guidal:2008ie}
%\bibitem{Guidal:2008ie}
  M.~Guidal,
  %``A fitter code for Deep Virtual Compton Scattering and Generalized Parton
  %Distributions,''
  Eur.\ Phys.\ J.\  A {\bf 37}, 319 (2008)
  [Erratum-ibid.\  A {\bf 40}, 119 (2009)]
  [arXiv:0807.2355 [hep-ph]].
  %%CITATION = EPHJA,A37,319;%% 

\bibitem{historyofDVCS}
D.~M{\"u}ller {\em et al.},
%``Wave functions, evolution equations and evolution kernels from
%light-ray operators of {QCD},''
Fortsch.\ Phys.\  {\bf 42}, 101 (1994);
%%CITATION = HEP-PH 9812448;%%

X.~Ji,
%``Gauge invariant decomposition of nucleon spin,''
Phys.\ Rev.\ Lett.\  {\bf 78}, 610 (1997);
%%CITATION = HEP-PH 9603249;%%

A.~V.~Radyushkin,
%``Nonforward parton distributions,''
Phys.\ Rev.\  {\bf D56}, 5524 (1997);
%%CITATION = HEP-PH 9704207;%%

J.~C.~Collins and A.~Freund,
%``Proof of factorization for deeply virtual Compton scattering in
%{QCD},''
Phys.\ Rev.\  {\bf D59}, 074009 (1999).
%%CITATION = HEP-PH 9801262;%%

\bibitem{gpdrev}
M.~Diehl,
  %``Generalized parton distributions,''
  Phys.\ Rept.\  {\bf 388} (2003) 41;
  %%CITATION = PRPLC,388,41;%%

A.~V.~Belitsky and A.~V.~Radyushkin,
  %``Unraveling hadron structure with generalized parton distributions,''
  Phys.\ Rept.\  {\bf 418}, 1 (2005);
  %%CITATION = PRPLC,418,1;%%

\bibitem{Anikin}
I.~V.~Anikin, B.~Pire, L.~Szymanowski, O.~V.~Teryaev and S.~Wallon,
  %``On BLM scale fixing in exclusive processes,''
  Eur.\ Phys.\ J.\  C {\bf 42}, 163 (2005).
  %%CITATION = EPHJA,C42,163;%%
\bibitem{JiO}
  X.~D.~Ji and J.~Osborne,
  %``One-loop corrections and all order factorization in deeply virtual  Compton
  %scattering,''
  Phys.\ Rev.\  D {\bf 58}, 094018 (1998)
  [arXiv:hep-ph/9801260].
  %%CITATION = PHRVA,D58,094018;%%
%\cite{Mankiewicz:1997bk}

%\cite{Ji:1997nk}
%\bibitem{Ji:1997nk}
  X.~D.~Ji and J.~Osborne,
  %``One-loop QCD corrections to deeply-virtual Compton scattering: The  parton
  %helicity-independent case,''
  Phys.\ Rev.\  D {\bf 57}, 1337 (1998)
  [arXiv:hep-ph/9707254].
  %%CITATION = PHRVA,D57,1337;%%

\bibitem{Mankiewicz}
  L.~Mankiewicz, G.~Piller, E.~Stein, M.~Vanttinen and T.~Weigl,
  %``NLO corrections to deeply-virtual Compton scattering,''
  Phys.\ Lett.\  B {\bf 425}, 186 (1998)
  [arXiv:hep-ph/9712251].
  %%CITATION = PHLTA,B425,186;%%
  

\bibitem{Belitsky:1997rh}
  A.~V.~Belitsky and D.~Mueller,
  %``Predictions from conformal algebra for the deeply virtual Compton
  %scattering,''
  Phys.\ Lett.\  B {\bf 417}, 129 (1998)
  [arXiv:hep-ph/9709379].
  %%CITATION = PHLTA,B417,129;%%
%\cite{Belitsky:1999sg}
\bibitem{Belitsky:1999sg}
  A.~V.~Belitsky, D.~Mueller, L.~Niedermeier and A.~Schafer,
  %``Deeply virtual Compton scattering in next-to-leading order,''
  Phys.\ Lett.\  B {\bf 474}, 163 (2000)
  [arXiv:hep-ph/9908337].
  %%CITATION = PHLTA,B474,163;%%

\bibitem{BDP}
E.~R.~Berger, M.~Diehl and B.~Pire,
  %``Timelike Compton scattering: Exclusive photoproduction of lepton pairs,''
  Eur.\ Phys.\ J.\  C {\bf 23}, 675 (2002).
  %%CITATION = EPHJA,C23,675;%%
%\cite{Pire:2008ea}

\bibitem{PSW}
  B.~Pire, L.~Szymanowski and J.~Wagner,
  %``Can one measure timelike Compton scattering at LHC ?,''
  Phys.\ Rev.\  D {\bf 79}, 014010 (2009), Nucl.\ Phys.\ Proc.\ Suppl.\  {\bf 179-180}, 232 (2008)
  and  Acta Phys.\ Polon.\ Supp.\  {\bf 2}, 373 (2009).
  %%CITATION = APPXA,2,373;%%
  %%CITATION = NUPHZ,179-180,232;%%
  %%CITATION = PHRVA,D79,014010;%%  
 

\bibitem{NadelTuronski:2009zz}
  P.~Nadel-Turonski, T.~Horn, Y.~Ilieva, F.~J.~Klein, R.~Paremuzyan and S.~Stepanyan,
  %``Timelike Compton scattering: A first look,''
  AIP Conf.\ Proc.\  {\bf 1182}, 843 (2009).
  %%CITATION = APCPC,1182,843;%%

\bibitem{DDVCS}
  M.~Guidal and M.~Vanderhaeghen,
  %``Double deeply virtual Compton scattering off the nucleon,''
  Phys.\ Rev.\ Lett.\  {\bf 90}, 012001 (2003)
  [arXiv:hep-ph/0208275].
  %%CITATION = PRLTA,90,012001;%%  
  
%\bibitem{BelitskyMueller}
A.~V.~Belitsky and D.~Mueller,
  %``Exclusive electroproduction of lepton pairs as a probe of nucleon
  %structure,''
  Phys.\ Rev.\ Lett.\  {\bf 90}, 022001 (2003)
  [arXiv:hep-ph/0210313].
  %%CITATION = PRLTA,90,022001;%%

  %\cite{Kopeliovich:2010xm}
%\bibitem{Kopeliovich:2010xm}
  B.~Z.~Kopeliovich, I.~Schmidt and M.~Siddikov,
  %``DDVCS on nucleons and nuclei,''
  Phys.\ Rev.\  D {\bf 82}, 014017 (2010)
  [arXiv:1005.4621 [hep-ph]].
  %%CITATION = PHRVA,D82,014017;%%

\bibitem{NLOphenoDVCS}
%\cite{Freund:2001rk}
%\bibitem{Freund:2001rk}
  A.~Freund and M.~F.~McDermott,
  %``A next-to-leading order QCD analysis of deeply virtual Compton  scattering
  %amplitudes,''
  Phys.\ Rev.\  D {\bf 65}, 074008 (2002)
  [arXiv:hep-ph/0106319].
  %%CITATION = PHRVA,D65,074008;%%

%\cite{Freund:2001hm}
%\bibitem{Freund:2001hm}
  A.~Freund and M.~F.~McDermott,
  %``A next-to-leading order analysis of deeply virtual Compton scattering,''
  Phys.\ Rev.\  D {\bf 65}, 091901 (2002)
  [arXiv:hep-ph/0106124].
  %%CITATION = PHRVA,D65,091901;%%

%\cite{Freund:2001hd}
%\bibitem{Freund:2001hd}
  A.~Freund and M.~McDermott,
  %``A detailed next-to-leading order QCD analysis of deeply virtual Compton
  %scattering observables,''
  Eur.\ Phys.\ J.\  C {\bf 23}, 651 (2002)
  [arXiv:hep-ph/0111472].
  %%CITATION = EPHJA,C23,651;%%
\bibitem{DIS}
  G. Altarelli, R.K. Ellis and G. Martinelli, Nucl. Phys. B157 (1979) 461.
  
\bibitem{Kfactor}
  R.~Stroynowski,
  %``Lepton Pair Production In Hadron Collisions,''
  Phys.\ Rept.\  {\bf 71}, 1 (1981) and references therein.
  %%CITATION = PRPLC,71,1;%%

\bibitem{BDPpi}
E.~R.~Berger, M.~Diehl and B.~Pire,
  %``Probing generalized parton distributions in pi N --> l+ l- N,''
  Phys.\ Lett.\  B {\bf 523}, 265 (2001).
  %%CITATION = PHLTA,B523,265;%%
\bibitem{TGDA} 
Z.~Lu and I.~Schmidt,
%``Pion pair production in e+ e- annihilation,''
  Phys.\ Rev.\  D {\bf 73}, 094021 (2006)
  [Erratum-ibid.\  D {\bf 75}, 099902 (2007)] and A.~Afanasev, S.~J.~Brodsky, C.~E.~Carlson and A.~Mukherjee,
  %``Timelike Virtual Compton Scattering from Electron-Positron Radiative
  %Annihilation,''
  Phys.\ Rev.\  D {\bf 81}, 034014 (2010).
  %%CITATION = PHRVA,D81,034014;%%
%%CITATION = PHRVA,D73,094021;%%

\bibitem{GDA}
M.~Diehl {\it et. al.}
  %``Exclusive production of pion pairs in gamma* gamma collisions at large
  %Q**2,''
  Phys.\ Rev.\  D {\bf 62}, 073014 (2000) and  Phys.\ Rev.\ Lett.\  {\bf 81}, 1782 (1998).
  %%CITATION = PRLTA,81,1782;%%
  %%CITATION = PHRVA,D62,073014;%%

\bibitem{TDA}
B.~Pire and L.~Szymanowski,
  Phys.\ Rev.\  D {\bf 71}, 111501 (2005);
  %%CITATION = PHRVA,D71,111501;%%
J.~P.~Lansberg, B.~Pire and L.~Szymanowski,
  Phys.\ Rev.\  D {\bf 73}, 074014 (2006) and Phys.\ Rev.\  D {\bf 75}, 074004 (2007)
  [Erratum-ibid.\  D {\bf 77}, 019902 (2008)]
 %%CITATION = PHRVA,D75,074004;%%
 %%CITATION = PHRVA,D73,074014;%%

\bibitem{TTDA}
 B.~Pire and L.~Szymanowski,
  Phys.\ Lett.\  B {\bf 622}, 83 (2005);
  %%CITATION = PHLTA,B622,83;%%
J.~P.~Lansberg, B.~Pire and L.~Szymanowski,
  %``Production of a pion in association with a high-Q2 dilepton pair in
  %antiproton-proton annihilation at GSI-FAIR,''
  Phys.\ Rev.\  D {\bf 76}, 111502 (2007).
  %%CITATION = PHRVA,D76,111502;%%
  
\end{thebibliography}
\end{document}